\documentclass[twocolumn]{aastex63}
\usepackage{bm,amsmath} % for bold math symbols
\usepackage{CJKutf8}
\graphicspath{{./}{figures/}}

\newcommand{\prospector}{\texttt{Prospector}}

\begin{document}

\title{A New Census of the $\bm{0.2 < z < 3.0}$ Universe, Part II: The Star-Forming Sequence}

%% Note that the corresponding author command and emails has to come
%% before everything else. Also place all the emails in the \email
%% command instead of using multiple \email calls.
\correspondingauthor{Joel Leja}
\email{joel.leja@psu.edu}

\author[0000-0001-6755-1315]{Joel Leja}
\affil{Department of Astronomy \& Astrophysics, The Pennsylvania State University, University Park, PA 16802, USA}
\affil{Institute for Computational \& Data Sciences, The Pennsylvania State University, University Park, PA, USA}
\affil{Institute for Gravitation and the Cosmos, The Pennsylvania State University, University Park, PA 16802, USA}

\author[0000-0003-2573-9832]{Joshua S. Speagle (\begin{CJK*}{UTF8}{gbsn}沈佳士\ignorespacesafterend\end{CJK*})}
\altaffiliation{Banting \& Dunlap Fellow}
\affiliation{Department of Statistical Sciences, University of Toronto, Toronto, ON M5S 3G3, Canada}
\affiliation{David A. Dunlap Department of Astronomy \& Astrophysics, University of Toronto, Toronto, ON M5S 3H4, Canada}
\affiliation{Dunlap Institute for Astronomy \& Astrophysics, University of Toronto, Toronto, ON M5S 3H4, Canada}

\author[0000-0001-5082-9536]{Yuan-Sen Ting (\begin{CJK*}{UTF8}{gbsn}丁源森\ignorespacesafterend\end{CJK*})}
\altaffiliation{Hubble Fellow}
\affiliation{Research School of Astronomy \& Astrophysics, Australian National University, Cotter Rd., Weston, ACT 2611, Australia}
\affiliation{Institute for Advanced Study, Princeton, NJ 08540, USA}
\affiliation{Department of Astrophysical Sciences, Princeton University, Princeton, NJ 08540, USA}
\affiliation{Observatories of the Carnegie Institution of Washington, 813 Santa Barbara Street, Pasadena, CA 91101, USA}
\affiliation{Research School of Computer Science, Australian National University, Acton ACT 2601, Australia}

\author[0000-0002-9280-7594]{Benjamin D. Johnson}
\affil{Center for Astrophysics\:$|$\:Harvard \& Smithsonian, 60 Garden St. Cambridge, MA 02138, USA}

\author[0000-0002-1590-8551]{Charlie Conroy}
\affil{Center for Astrophysics\:$|$\:Harvard \& Smithsonian, 60 Garden St. Cambridge, MA 02138, USA}

\author[0000-0001-7160-3632]{Katherine E. Whitaker}
\affil{Department of Astronomy, University of Massachusetts Amherst, 710 N Pleasant Street, Amherst, MA 01003, USA}
\affil{Cosmic Dawn Center (DAWN), Copenhagen, Denmark}

\author[0000-0002-7524-374X]{Erica J. Nelson}
\affil{Astrophysical \& Planetary Sciences, 391 UCB, 2000 Colorado Ave, Boulder, CO 80309, Duane Physics Building, Rm.
E226}

\author[0000-0002-8282-9888]{Pieter van Dokkum}
\affil{Department of Astronomy, Yale University, New Haven, CT 06511, USA}

\author[0000-0002-8871-3026]{Marijn Franx}
\affil{Leiden Observatory, Leiden University, NL-2300 RA Leiden, Netherlands}

\submitjournal{ApJ}
\begin{abstract}
We use the panchromatic SED-fitting code \texttt{Prospector} to measure the galaxy logM$^*$-logSFR relationship (the `star-forming sequence') across $0.2 < z < 3.0$ using the COSMOS-2015 and 3D-HST UV-IR photometric catalogs. We demonstrate that the chosen method of identifying star-forming galaxies introduces a systematic uncertainty in the inferred normalization and width of the star-forming sequence, peaking for massive galaxies at $\sim 0.5$ dex and $\sim0.2$ dex respectively. To avoid this systematic, we instead parameterize the density of the full galaxy population in the logM$^*$-logSFR-redshift plane using a flexible neural network known as a normalizing flow. The resulting star-forming sequence has a low-mass slope near unity and a much flatter slope at higher masses, with a normalization $0.2-0.5$ dex lower than typical inferences in the literature. We show this difference is due to the sophistication of the \texttt{Prospector} stellar populations modeling: the nonparametric star formation histories naturally produce higher masses while the combination of individualized metallicity, dust, and star formation history constraints produce lower star formation rates than typical UV+IR formulae. We introduce a simple formalism to understand the difference between SFRs inferred from spectral energy distribution fitting and standard template-based approaches such as UV+IR SFRs. Finally, we demonstrate the inferred star-forming sequence is consistent with predictions from theoretical models of galaxy formation, resolving a long-standing $\sim0.2-0.5$ dex offset with observations at $0.5<z<3$. The fully trained normalizing flow including a nonparametric description of $\rho(\log{\rm M}^*,\log{\rm SFR},z)$ is made available online \footnote{\url{https://github.com/jrleja/sfs_leja_trained_flow}} to facilitate straightforward comparisons with future work.
\end{abstract}

\section{Introduction}
The star-forming sequence is the relationship between galaxy star formation rate (SFR) and stellar mass (M$^*$) observed in star-forming galaxies.\footnote{This is often called the ``main sequence'' of star-forming galaxies in the literature, although we do not use the term here to avoid confusion with the stellar main sequence.} The majority of galaxies form the majority of their mass either ``on'' (e.g. \citealt{leitner12}) or ``passing through'' (e.g. \citealt{abramson15}) the star-forming sequence, making a robust characterization of the star-forming sequence crucial to understanding the process of galaxy evolution as a whole. As a result, the star-forming sequence has been well-measured in the literature across a variety of rest-frame wavelengths and with a range of assumptions and techniques \citep{daddi07,noeske07,karim11,rodighiero11,whitaker12,whitaker14,speagle14,renzini15,schreiber15,tomczak16,leslie20}.

Yet despite the importance of the star-forming sequence, there is a large dispersion across the literature in its inferred normalization and slope. \citet{speagle14} performed an exhaustive cross-calibration of measurements of the star-forming sequence in the astronomical literature. They showed that there are two key sets of systematics which drive the dispersion in measurements of the star-forming sequence. 

The first systematic is the method chosen to identify star-forming galaxies and -- usually at the same time -- to separate quiescent galaxies. These different methods include rest-frame color-color selection (e.g., UVJ \citealt{williams09}, sBzK \citealt{daddi04}), fixed lines in the SFR-M$_*$ plane (e.g., specific star formation rate cuts), lower limits on observed emission line luminosities or equivalent widths, $\sigma$-clipping fits to the star-forming sequence, Lyman-break or luminous infrared galaxy (LIRG) selection, or no pre-selection at all. \citet{speagle14} demonstrates that the measured slope of the log(SFR)-log(M$^*$) relationship ranges between $0-1$ in the literature, and this variance correlates strongly with the chosen method of identifying and classifying star-forming galaxies. This result has also been observed in simulations. For instance, \citet{donnari19} contrasts rest-frame color-based selection with selection based on bimodality in the SFR-mass plane in the Illustris-TNG hydrodynamical simulation of galaxy formation. They find that this choice affects the normalization of the star-forming sequence at the $\sim0.5$ dex level, and while these selections generally produce identical fractions of quiescent galaxies at low stellar masses (to within 5-10\%), this measured fraction will vary between 20-40\% at high masses (M/M$_{\odot} > 10^{10.5}$).

The second systematic challenge in observational measurements of the star-forming sequence is the method by which galaxy star formation rates and (to a somewhat lesser extent) stellar masses are inferred. \citet{speagle14} finds an interpublication 1$\sigma$ scatter of $\sim0.2$ dex in the measured normalization of the star-forming main sequence, the great majority of which \citet{speagle14} attributes to systematics in the inference techniques for galaxy SFRs and stellar masses. Again, these challenges are straightforward to reproduce in simulated data: \citet{katsianis20} calculate UV, IR, and spectral energy distribution (SED) SFRs from observations of simulated galaxies including dusty radiative transfer effects and find even more dramatic offsets of $0.3-0.5$ dex, with additional strong systematic trends with redshift.

By creating calibration offsets associated with each measurement methodology, \citet{speagle14} is able to bring these independent observational measurements of the star-forming sequence in the literature onto a common scale. This shows that these offsets are {\it precise} and {\it predictable}; however, it does not ensure that one or any of them are {\it accurate}.

More insight can be had by examining the star-forming sequence in simulations of galaxy formation. These simulations produce star-forming sequences which are systematically $0.2-0.5$ dex lower than the observed star-forming sequences at $z=0.5-3$, yet are still able to reproduce the evolution of the stellar mass function with reasonable accuracy. This offset has remained for nearly a decade and is present across a diverse range of galaxy formation models such as Illustris, both the original \citep{sparre15} and TNG \citep{donnari19}; SIMBA \citep{dave19}; EAGLE \citep{furlong15}; the Santa Cruz semi-analytical model \citep{somerville15}; and other analytical models of galaxy formation \citep{lilly13,dekel14}. Indeed, matching these observations seems to require a strong but temporary decoupling of baryonic growth from halo growth at $1 < z < 3$ \citep{mitchell14}. Indeed, direct calculations suggest the simulations may be attempting to match two incompatible sets of observations, as the total mass growth implied by the observed evolution of the mass function is lower than the total star formation rate density inferred from the observed star-forming sequence by $\sim0.3$ dex \citep{leja15}.

It is therefore crucial to develop better machinery to measure the star-forming sequence from observations, both to better characterize the star-forming sequence itself -- which plays a central role in galaxy evolution -- and to provide more accurate constraints on this sequence to simulations. One potential solution to the challenge of quiescent/star-forming separation is the adoption of data-driven definitions of the star-forming sequence which do not rely on any {\it a priori} separation between star-forming and quiescent galaxies. Recent solutions in the literature include defining the star-forming sequence as the ``ridgeline'' (i.e. the density peak in the SFR-mass plane) \citep{renzini15}, or using flexible Gaussian mixture models with the number of Gaussian sub-components determined by the data \citep{hahn19}. In the limit of wide-area surveys, these data-driven approaches can accurately measure higher-order moments of the star-forming sequence such as the width or the location of the so-called `green valley` \citep{sherman21}.

There are also methodologies emerging to infer stellar populations parameters from observations more robustly, producing a promising way forward to mitigating or even eliminating the systematics in SFR and mass measurements. Often SFRs are inferred using direct flux-to-SFR conversions which are derived via fixed assumptions about the properties of galaxy stellar populations (e.g. \citealt{kennicutt12}). This can result in significant measurement systematics when applied to a galaxy population with a rich diversity of properties \citep{wuyts11a,leja19b}. New generations of Bayesian galaxy SED-models using highly flexible on-the-fly model generation (e.g., \texttt{Beagle}, \citealt{chevallard16}; \texttt{Prospector}, \citealt{leja17,johnson21}) or large pre-generated grids of models (e.g. \texttt{Bagpipes}, \citealt{carnall18}; \texttt{CIGALE}, \citealt{boquien19}) can now fit sophisticated models to all of the available data. These methods constrain stellar populations properties such as star formation histories, dust attenuation, and metallicities on an object-by-object basis, as opposed to earlier works using fixed values or highly simplified models. The net effect is to produce more sophisticated (and, ideally, more accurate) conversions between the observed flux and the star formation rate and stellar masses.

In this work, we measure the star-forming sequence using stellar populations properties inferred by the \texttt{Prospector-$\alpha$} model built in the \prospector{} SED-fitting code. This work is motivated by the distinct differences in inferred masses and star formation rates from previous analyses \citep{leja19b}. We additionally use a new machine learning technique to measure the star-forming sequence: normalizing flows \citep{jimenez15}. Normalizing flows allow the direct inference of the density in the SFR-M$^*$ plane from individual observations. Crucially, this density inference can be modified to marginalize over measurement uncertainty as described later in this work -- indeed, \citet{curtis-lake21} has shown that proper modeling of the measurement uncertainties has a significant effect on the inferred slope and normalization of the star-forming sequence. This density estimate in turn permits measurement of the `ridge' of the star-forming sequence without any pre-selection of star-forming galaxies, as well as facilitating direct comparisons with simulations or other observations using {\it density} in M$^*$-SFR-redshift space. This paper is the second in a series; the first paper presented a Bayesian population model for the stellar mass function \citep{leja20}.

The paper is structured as follows. Section \ref{sec:data} describes the photometric data and the subsequent SED-modeling. Section \ref{sec:bimodality} uses these data to demonstrate the sensitivity of the star-forming sequence to a bimodal classification system. Section \ref{sec:nf} introduces the normalizing flow method and describes our alterations to the standard methodology to include marginalization over the parameter uncertainties from SED modeling. Section \ref{sec:results} analyzes the resulting star-forming sequence and presents simple equations describing the evolution with stellar mass and redshift. Section \ref{sec:comparison} compares to results in the literature, and provides a general framework in which to understand how SED-based SFRs are different than standard UV+IR formulae. Section \ref{sec:discussion} contains a broader discussion of the results and the conclusion is in Section \ref{sec:conclusion}. We use a \citet{chabrier03} initial mass function and adopt a WMAP9 cosmology \citep{hinshaw13} with $H_0=69.7$ km/s/Mpc, $\Omega_b = 0.0464$, and $\Omega_c=0.235$. Reported star formation rates are averaged over the most recent 100 Myr (rather than instantaneous values) unless explicitly stated otherwise.

\section{Data and SED Modeling}
\label{sec:data}
Here we describe how the inputs to the star-forming sequence (i.e., stellar masses, star formation rates, and redshifts) are derived. The input photometry and SED fitting methodology are identical to Paper I \citep{leja20} and are briefly summarized here for convenience. The first two sections describe the publicly-available surveys from which photometry and redshifts are taken. The subsequent section describes the SED modeling used to translate these data into stellar population properties.

\subsection{3D-HST Catalog}
\label{sec:3dhst}
The 3D-HST photometric catalogs \citep{skelton14} collate all available public photometry in five deep, well-studied extragalactic fields comprising an area of $\sim$900 arcmin$^2$. In particular, this includes deep near-IR imaging from the CANDELS HST Treasury Program \citep{grogin11,koekemoer11}. This photometric catalog is supplemented with $Spitzer$/MIPS 24$\mu$m photometry from \citet{whitaker14}. In total, each galaxy is covered over 0.3-24$\mu$m in the observed-frame with between 17-44 bands of aperture-matched photometry.

The redshifts come from several sources. The 3D-HST grism survey \citep{momcheva16} provides reliable space-based grism redshifts. The rest of the sample consists of photometric redshifts estimated with the publicly-available EAZY \citep{brammer08}. Approximately 30\% of the sample redshifts are from ground-based spectroscopy or reliable grism measurements, while the remainder are based on photometric estimates. The photometric redshifts are taken as the peak of the redshift probability distribution function. Thanks to the wide wavelength coverage and mixture of narrow-band and medium-band filters, they show a relatively low scatter of 0.0197$(1+z)$ compared to the grism redshifts \citep{bezanson16}.

We select a stellar mass-complete sample of 26,971 galaxies at $0.5 < z < 3.0$ from this catalog. The lower redshift limit is set by the redshift at which the aperture-based photometry measurements start to become unreliable, and the upper redshift limit is where space-based (i.e. $HST$) rest-frame optical imaging is no longer available for target selection. Mass-complete limits are taken from \citet{leja20}, with an adjustment upwards by 0.1 dex in this work at $z=0.65$ and $z=2.1$ to remove an observed upturn in the star-forming sequence at the lowest masses. This is necessary because the stellar mass-complete limit is set at 90\%, and a 10\% incompleteness rate can have a substantive effect on $<$SFR$>$, as red galaxies with low SFRs are more likely to drop out. The level of this adjustment is determined by examining the log$M^*$ - logSFR distribution in small redshift slices at each redshift, and requiring that the distribution of galaxies in logSFR at the stellar mass-complete limit is similar to that at slightly higher stellar masses. As none of the conclusions are sensitive to the SFR near the mass-complete limit, this adjustment has no effect on the conclusions of this work.

\subsection{COSMOS-2015 Catalog}
We additionally include photometry and redshifts from the COSMOS-2015 catalog \citep{laigle16} to bolster the low-redshift volume. This includes 30 bands of photometry covering the observed-frame $0.2-24\mu$m, with a small fraction ($<1\%$) of objects with $Herschel$ far-IR imaging. We select only objects in the overlap with the UltraVISTA survey \citep{mccracken12} to include crucial near-IR photometric bands, producing an effective area of 1.38 deg$^2$.

The majority of the redshifts are photometric, estimated using LePhare \citep{ilbert06}. These have a very low scatter of 0.007-0.009$(1+z)$ in the targeted redshift range \citep{laigle16}, though notably this comparison is only performed for bright galaxies targeted by spectroscopic programs.

We select a stellar-mass complete sample of 41,148 objects at $0.2 < z < 0.8$ from this catalog, where the lower limit avoids the saturation limit for bright galaxies \citep{davidzon17} and the upper limit ensures overlap with the 3D-HST sample. The mass-complete limits are again taken from \citet{leja20}.  Following the same procedure described in Section \ref{sec:3dhst}, we also adjust the $z=0.175$ ($z=0.5$) mass-complete limit in COSMOS-2015 upwards by 0.25 (0.1) dex.

\subsection{SED Modeling}
The photometry is fit using the galaxy SED-fitting code \prospector{} \citep{leja17,johnson21}, powered by the FSPS stellar population synthesis code \citep{conroy09b}. Within FSPS, the MIST stellar isochrones are adopted \citep{choi16,dotter16}.

The physical model is the \texttt{Prospector-$\alpha$} model from \citet{leja19b}, which is a modified version of the model from \citet{leja17}. In brief, the model has 14 free parameters, consisting of a 7-component nonparametric star formation history using the `continuity' prior which disfavors sharp changes in SFR(t) \citep{leja19a}, a two-component dust attenuation model with a flexible dust attenuation curve \citep{noll09}, free gas-phase and stellar metallicity, and a mid-infrared AGN component with a free normalization and dust optical depth. \prospector{} includes dust emission powered by energy balance \citep{dacunha08} with an SED  and nebular emission self-consistently powered by the model stellar ionizing continuum \citep{byler17}.

%An important motivation for this analysis is that \texttt{Prospector-$\alpha$} produces star formation rates and stellar masses which are offset from standard UV+IR SFRs and stellar masses by $\sim0.1-0.3+$ dex \citep{leja19b}. This is discussed in detail in Section XXX.

Notably, in contrast to typical UV+IR SFRs, star formation rates can be measured via this methodology for every object in the survey regardless of whether they are detected in $Spitzer$/MIPS 24$\mu$m. Of course, objects without IR detections will naturally have larger parameter uncertainties.

\begin{figure*}
\begin{center}
\includegraphics[width=1.0\linewidth]{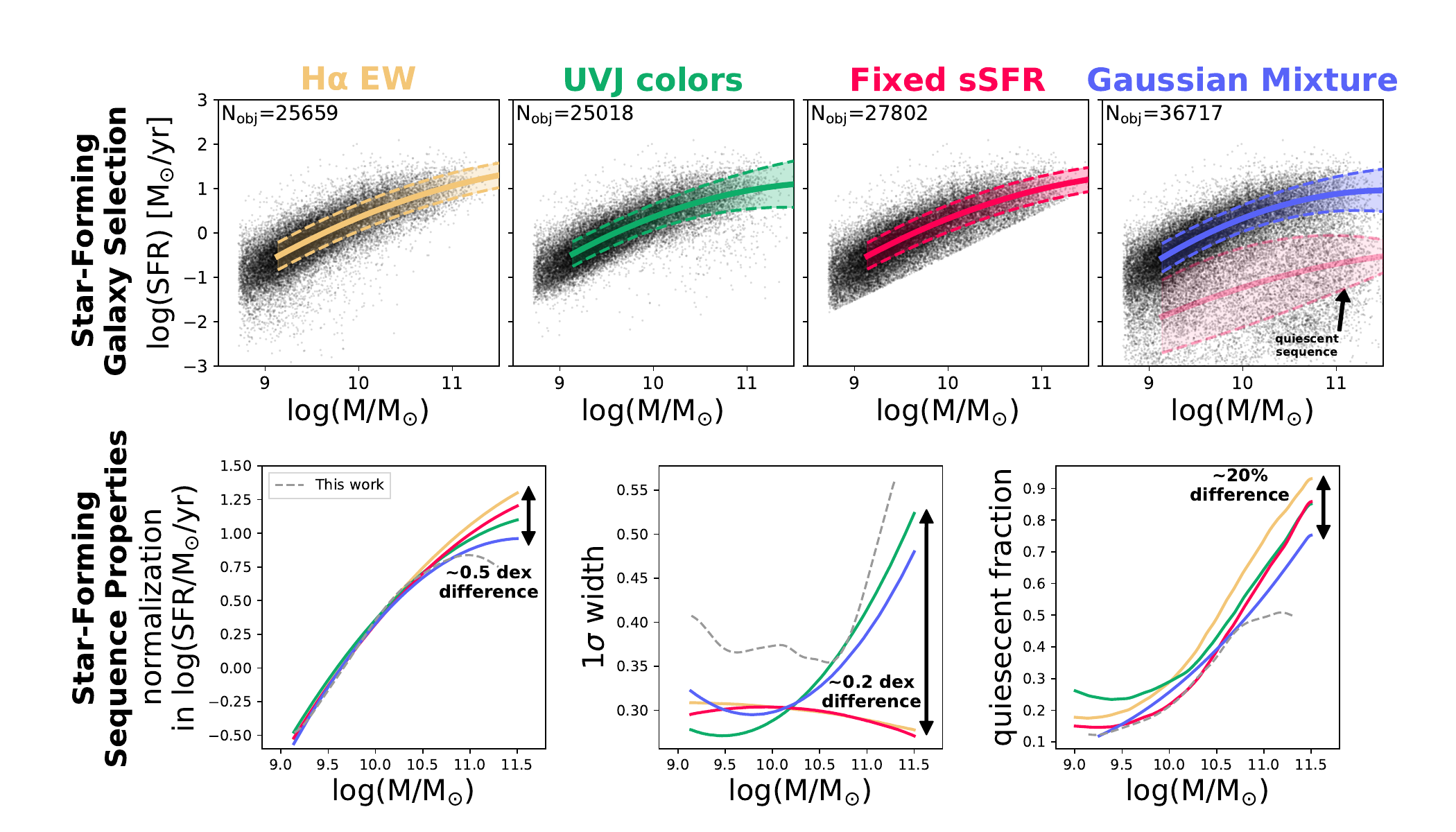}
\caption{Four different methods to select star-forming galaxies, and the properties of the resulting star-forming sequences. The top panels show the M$^*$-SFR plane for galaxies at $0.3< z <0.7$ identified as star-forming via four different techniques, with the total number of objects included in the fit indicated in the upper-left and the mean and $1\sigma$ scatter of the resulting star-forming sequence as solid and dashed lines respectively. For the Gaussian mixture model, the quiescent sequence is also shown. The lower panels show the median and $1\sigma$ width of the resulting star-forming sequence and the fraction of quiescent galaxies inferred from each method. For reference, we additionally show the values derived later in this paper with a grey dashed line.}
\label{fig:bimodal_split}
\end{center}
\end{figure*}

\section{Demonstrating the Sensitivity to Bimodal Classification}
\label{sec:bimodality}
Here we use the SED-fitting posteriors from the previous section to demonstrate the sensitivity of the star-forming sequence to the method adopted for identifying star-forming galaxies. We select galaxies between $0.3 < z < 0.7$ from our data, separate them into star-forming and quiescent galaxies using four different methods described below, and fit a Gaussian with a mass-dependent mean and width to the distribution of log(SFR) in the identified star-forming galaxies. The mean and standard deviation of the Gaussian are allowed to vary quadratically with stellar mass. We take $N=50$ posterior samples for each galaxy and marginalize the likelihood over the parameter uncertainties, following \citet{leja20}. This effectively results in an error-deconvolved model of the star-forming sequence. We explore the posterior with the nested sampling code \texttt{dynesty} \citep{speagle20}.

We examine four common methods to separate between star-forming and quiescent galaxies:

\begin{enumerate}
    \item {\bf H$\mathbf{\alpha}$ equivalent width}: The H$\alpha$ equivalent width is estimated using the \prospector{} posterior predictions for each galaxy. These predictions are reasonably accurate, showing a $1\sigma$ scatter of $\sim 0.2$ dex \citep{leja17} when compared to spectroscopic observations. We use an EW cut of 35\AA{}, which is approximately $5-10$\AA\ below the mean equivalent width of star-forming galaxies at this epoch \citep{fumagalli12}.
    \item {\bf UVJ colors}: Rest-frame UVJ colors are synthesized from the \prospector{} posteriors for each galaxy, and the galaxy is assigned to be star-forming or quiescent based on the UVJ quiescent region from \citet{whitaker12}. When the object straddles the quiescent/star-forming boundary, the classification is assigned following the classification of the {\it majority} of the posterior weight.
    \item {\bf Fixed sSFR}: A fixed sSFR boundary at log(sSFR/yr$^{-1}) = -10.5$ separates star-forming and quiescent galaxies, with objects near the border again assigned based on the majority of their posterior weight.
    \item {\bf Gaussian Mixture}: Two Gaussians are fit to the log(SFR)-log(M$^*$) distribution, where here the second Gaussian is intended to describe the quiescent sequence. The means and widths vary quadratically with mass as before. To ensure robust separation between these two sequences, the mean of the second Gaussian is defined relative to the first one via:
    \begin{equation}
    x_{\mathrm{qu}} = x_{\mathrm{sf}} - (\sigma_{\mathrm{SF}} + \sigma_{\mathrm{QU}}) - \Delta_{\mathrm{QU}}
    \end{equation}
    where $\Delta_{\mathrm{QU}}$ is the fit parameter. This parameter has a uniform prior taken between 0.0 and 2.0, $\sigma_{\mathrm{SF}}$ and $\sigma_{\mathrm{qu}}$ are the standard deviations of the star-forming and quiescent sequences, and $x_{\mathrm{sf}}$ and $x_{\mathrm{qu}}$ are their means. This formulation ensures that the two sequences have a minimum separation set by the sum of their standard deviations.
\end{enumerate}

The results of these fits are shown in Figure \ref{fig:bimodal_split}. There is reasonable agreement between the techniques below log(M/M$_{\odot})\lesssim$10.5, with agreement in the normalization of the star-forming sequence to within $\lesssim0.1$ dex and to within $0.05$ dex for the $1\sigma$ width.

The striking differences occur at the massive end, where there is both a larger percentage of quiescent galaxies and a much less clear delineation between the two populations in the logM$^*$-logSFR plane (both of these effects can be seen directly in the Gaussian mixture panel of Figure \ref{fig:bimodal_split}). As a result, there is a $0.5$ dex range in the inferred normalization at the massive end and a $0.2$ dex range in the inferred $1\sigma$ widths. UVJ colors in particular produce a wide star-forming sequence at the massive end, a result that naturally follows from the flattening of the relationship between UVJ colors and specific star formation rate at log(sSFR/yr$^{-1})<-10.5$ \citep{leja19c}.

The inferred quiescent fraction shows substantial variation, with a difference of $\sim20$\% between techniques that shows no clear dependence on stellar mass (see \citealt{donnari19} for similar conclusions for simulated galaxies). Interestingly, no one technique produces a consistently higher or lower quiescent fraction at all masses, demonstrating how the traditional observational signatures of quiescence are also affected by natural variation in underlying stellar populations properties (e.g., the level of time-variability in star formation histories will have a different effect on H$\alpha$ vs $UVJ$ quiescent indicators).

We also include the results from the analysis of the full population described later in this paper as grey dashed lines. These results are not quite directly comparable as the analysis is performed on the full population, but are useful as a point of comparison.

\section{Inferring the Population Density With Normalizing Flows}
\label{sec:nf}

\begin{figure*}[ht!]
\begin{center}
\includegraphics[width=0.95\linewidth]{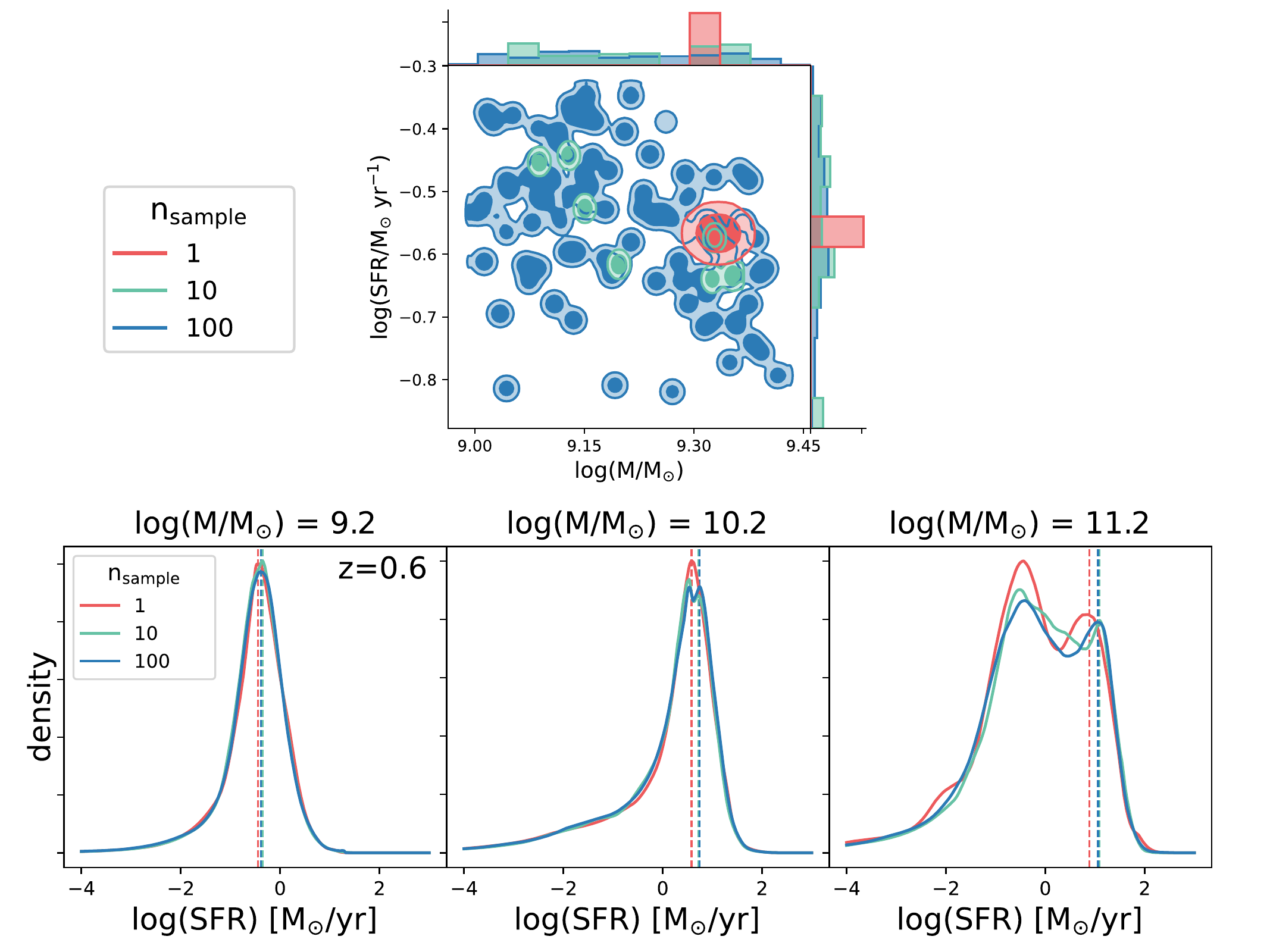}
\caption{The effect of incorporating uncertainties in logM$^*$ and logSFR on the inferred density, P(logSFR$|$logM$^*,z$). The top panel shows the joint log\,M$^*$-log\,SFR posterior for a single object, demonstrating that increasing the number of posterior samples for one object produces a better resolution in both the 1D uncertainties for mass and star formation rate, and of the correlation between them. The lower panels show the density profile across the star-forming sequence at fixed M$^*$ and redshift. The colors in each panel denotes how many posterior samples were included from each galaxy when inferring the density in (log\,M$^*$, log\,SFR, z) with the normalizing flow method. The ridge of the star-forming sequence is marked with a dashed line. Including uncertainties has a relatively strong effect on the profile of the sequence at high masses, shifting the median by $\sim0.2$ dex. This sensitivity arises because the uncertainty on the average star formation rate is relatively high due to the larger number of low sSFR systems, and the number of objects is relatively low. In contrast, measurement uncertainty has a relatively negligible effect on lower mass systems, likely due to the fact that they are far more numerous due to the shape of the galaxy stellar mass function.}
\label{fig:nsamp_comparison}
\end{center}
\end{figure*}

In light of the challenges posed in cleanly separating star-forming and quiescent galaxies, we instead proceed by inferring the density $\rho(\log \mathrm{M}^*, \log {\rm SFR}, z)$ in stellar mass, SFR, and redshift for the full galaxy population using a normalizing flow. From this, the star-forming sequence can be inferred directly by tracing the peak in the distribution of log\,SFR as a function of stellar mass and redshift, following \citet{renzini15}. We provide a brief description of normalizing flows below, along with some additional implementation details. It is not necessary to fully understand the normalizing flow treatment to follow the results of the paper, though it is important to note that the implementation below marginalizes over the physical parameter uncertainties for each galaxy (see Fig. \ref{fig:nsamp_comparison}).

A normalizing flow is a type of neural network that tries to learn a non-parametric estimate of the continuous, underlying probability density function (PDF) $P_{\theta}(\mathbf{x})$ associated with a set of $N$ observations $\{ \mathbf{x}_1, \dots, \mathbf{x}_N \}$ given some parameters $\theta$ which describe the transformation into the target distribution. For our purposes, $\mathbf{x} = \{ \log {\rm M}^*, \log {\rm SFR}, z\}$. The basic method by which a normalizing flow works is to determine a series of transformations $f_{\theta_k}$ from latent variables $\mathbf{z}$ (distinct from redshift, $z$) with a known simple distribution $P(\mathbf{z})$ to the observations $\mathbf{x}$ with unknown distribution $P_\theta(\mathbf{x})$. Here, $\theta$ is the set of parameters in the neural network used to transform to the target distribution. In most cases, $\mathbf{z}$ is assumed to follow a set of iid (independent and identically distributed) standard Normal (i.e. unit Gaussian) distributions. In other words, the normalizing flow is trying to learn a series of transformations from three unit Gaussians to reproduce the distribution of the data in $\log {\rm M}^*$, $\log {\rm SFR}$, and $z$. 

Assuming that the data follow a point process and are independent and identically distributed (iid), a normalizing flow works by selecting a given $\theta_{\rm max}$ that maximizes the log-likelihood
\begin{equation}
    \theta_{\rm max} = {\rm argmax}_{\theta} \left( \sum_{i=1}^{N} \log P_{\theta}(\mathbf{x}_i) \right)
\end{equation}

Taking advantage of the change of variables formula, for a series of $k$ composite transformations
\begin{equation}
    \mathbf{x} = f_\theta(\mathbf{z}) = f_{\theta_k}(\cdots f_{\theta_2}(f_{\theta_1}(\mathbf{z})))
\end{equation}
where we have split $\theta = \{ \theta_1, \dots, \theta_k\}$ into $k$ portions, the PDF at any particular point depends on the initial PDF multiplied by the Jacobian:
\begin{equation}
    P(\mathbf{x}) = P(f_\theta(\mathbf{z})) = P(\mathbf{z}) \times \left| \frac{\partial f_\theta(\mathbf{z})}{\partial \mathbf{z}} \right|^{-1}
\end{equation}
Exploiting the chain rule and taking the logarithm, we can rewrite this expression as
\begin{equation}
    \log P(\mathbf{x}) = \log P(f_\theta(\mathbf{z})) = \log P(\mathbf{z}) - \sum_{j=1}^{k} \log \left| \frac{\partial f_{\theta_j}(\mathbf{z}^{j-1})}{\partial \mathbf{z}^{j-1}} \right|
\end{equation}
where $\mathbf{z}^j = f_{\theta_j}(\cdots f_{\theta_1}(\mathbf{z}))$ and $\mathbf{z}^0 = \mathbf{z}$.
The core challenge for training normalizing flows is to ensure that the series of transformations $f_{\theta_j}$ have easy-to-compute Jacobians (to evaluate the log-PDF) and inverses $f^{-1}_{\theta_j}$ (to go from $\mathbf{x} \rightarrow \mathbf{z}$ as well as $\mathbf{z} \rightarrow \mathbf{x}$) while still retaining the flexibility to model complex distributions.

For the normalizing flow in this work, we build on the same basic architecture and training procedure as \citet{ting21}, using a batch size of 1024 and 500 epochs. This architecture consists of eight units of a ``Neural Spline Flow'', each of which consists of three layers of densely connected neural networks with 16 neurons, coupled with a ``Conv1x1'' operation (often referred to as ``GLOW'' in the machine learning literature; see, e.g., \citealt{kingma18,durkan19}). See \citet{green20,ting21} for additional details, particularly Fig. 3 of \citet{ting21} which visualizes the efficacy of these transformations using complex analytical PDFs.

We elaborate on one feature of the above approach. Due to the coupling between neural network layers, the normalizing flow does not perform well with a small number of odd parameters (three in this work). As a result, we introduce a dummy variable $\epsilon$ drawn from a standard normal distribution and train our normalizing flow to infer $P_\theta(\log \mathrm{M}^*, \log {\rm SFR}, z, \epsilon)$. We subsequently marginalize over $\epsilon$ when calculating the final PDF $P_\theta(\log \mathrm{M}^*, \log {\rm SFR}, z)$.

We also make a modification to the above approach. This modification takes into account that the quantities inferred from the observations -- $\log \mathrm{M}^*$, $\log {\rm SFR}$, and $z$ -- are actually noisy and probabilistic estimates of their true underlying values $\log {\rm M}^*_{\rm true}$, $\log {\rm SFR}_{\rm true}$, and $z_{\rm true}$. Computing the log-probability for each observation for our normalizing flow therefore requires marginalizing over the corresponding uncertainty $P_{\theta}(\mathbf{x}_i | \mathbf{x}_{i, {\rm true}})$:
\begin{equation}
    P_{\theta}(\mathbf{x}_i) = \int P_{\theta}(\mathbf{x}_{i, {\rm true}}) P(\mathbf{x}_i | \mathbf{x}_{i, {\rm true}}) \,{\rm d}\mathbf{x}_{i, {\rm true}}
\end{equation}
Following the strategy in \citet{leja20}, we approximate this integral using a weighted average based on $M=100$ samples from the posterior distribution for each object:
\begin{equation}
    \int P_{\theta}(\mathbf{x}_{i, {\rm true}}) P(\mathbf{x}_i | \mathbf{x}_{i, {\rm true}}) \,{\rm d}\mathbf{x}_{i, {\rm true}} \approx \frac{\sum_{j=1}^{M} w_{ij} P_{\theta}(\mathbf{x}_{ij, {\rm true}})}{\sum_{j=1}^{M} w_{ij}}
\end{equation}
where $w_{i,j}$ are importance weights that mitigate the impact of the prior on $\log \mathrm{M}^*$, $\log {\rm SFR}$, and $z$. This effectively results in an uncertainty-deconvolved measurement of the SFR-stellar mass plane at a fixed redshift. We note that this only marginalizes over the uncertainties in free parameters in the \prospector-$\alpha$ physical model. It does not include more fundamental uncertainties (e.g., the IMF).

Figure \ref{fig:nsamp_comparison} shows the effect of explicitly accounting for the uncertainties this way, highlighting both the shape of the posterior in mass-SFR space for an individual galaxy and the vertical profile in the SFR-M plane at several slices. The error deconvolution has the greatest effect where the uncertainties are highest, namely at low specific star formation rates. Broadly speaking, the width of the star-forming sequence is not very sensitive to the error resolution except for this high-mass regime.

The subsequent analysis uses a fine sampling of the density field from the trained normalizing flow and derives the properties of the star-forming sequence from this sampled density field. We also smooth with a Gaussian kernel with $\delta z$ = 0.1 in redshift to account for the effects of sampling variance (cosmic variance) caused by to the finite areal coverage of the surveys. This is an imperfect solution; ideally, in future work it will be possible to train a normalizing flow which marginalizes over density fluctuations using an explicit model for cosmic sampling variance (e.g., \citealt{leja20}).

For convenience, in the rest of this work we will drop the $\theta$ subscript and instead refer to the actual 3-D density estimated by the flow
\begin{equation}
    \rho(\log \mathrm{M}^*, \log {\rm SFR}, z) = N \times P_{\theta}(\log \mathrm{M}^*, \log {\rm SFR}, z)
\end{equation}
where $N$ is the number of objects (galaxies) in our dataset.

\section{Results from Analysis of the logM$^*$-logSFR-z Density Field}
\label{sec:results}
Here we explore the density field $\rho(\log \mathrm{M}^*, \log {\rm SFR}, z)$ inferred by our normalizing flow. We calculate and parameterize the ridge and mean of the star-forming sequence in Section \ref{sec:sfs_results} and fit the profile of the star-forming sequence to derive the width and quiescent fraction in Section \ref{sec:derived_pars}.

\subsection{The star-forming sequence}
\label{sec:sfs_results}

\begin{figure*}[ht]
\begin{center}
\includegraphics[width=0.95\linewidth]{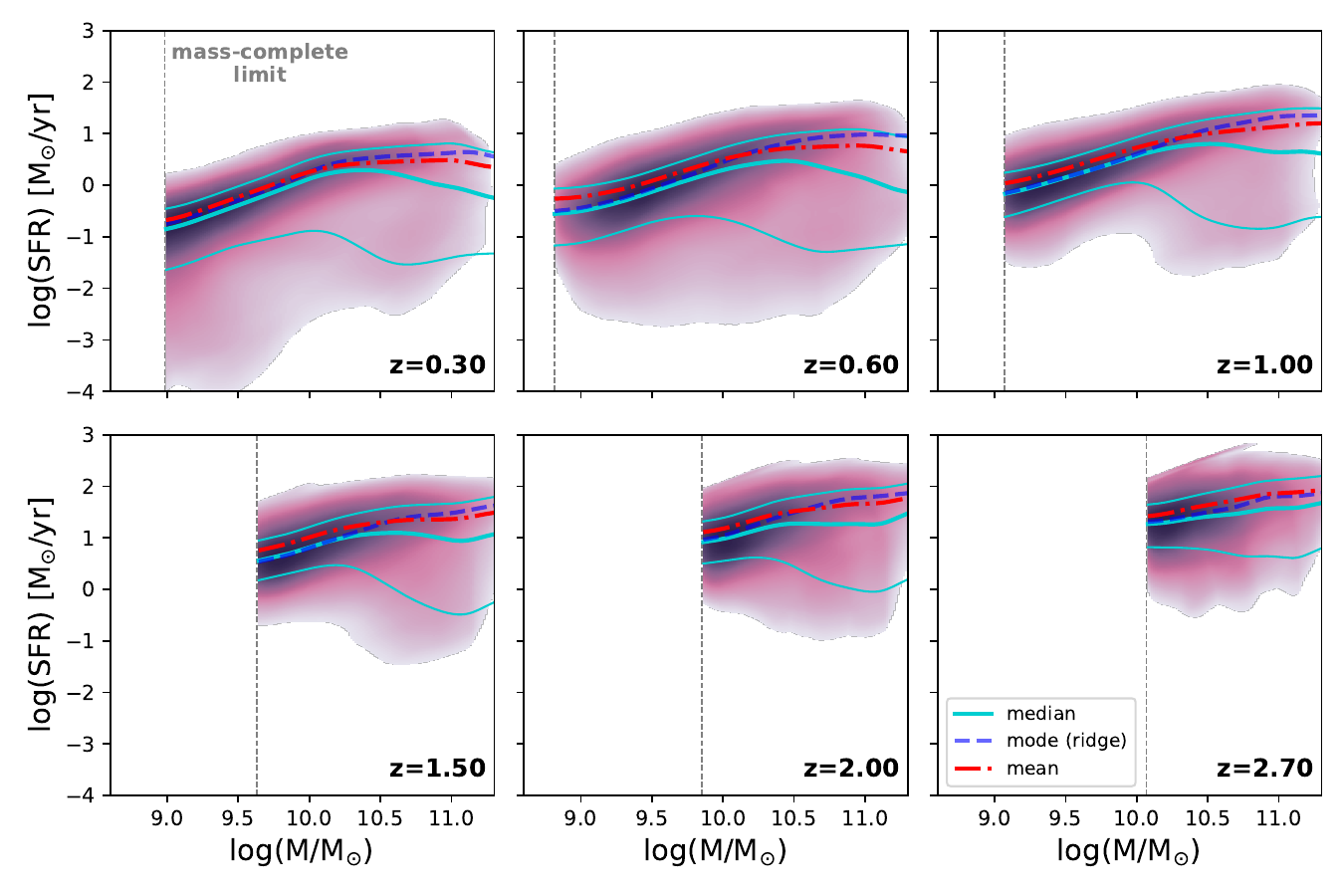}
\caption{The log\,M$^*$-log\,SFR relationship at six redshifts. The color shading indicates the log-density $\log \rho(\log \mathrm{M}^*, \log {\rm SFR}, z)$ estimated from our normalizing flow. The thick teal line show the running median, with the thin teal lines showing the 16th/84th percentiles. The dashed blue line shows the mode (i.e. the ridge-line), while the dash-dot red line shows the mean SFR (as opposed to mean logSFR). Below $z=0.5$, the mass-complete limit is from COSMOS-2015, while above $z=0.5$ it is from the deeper 3D-HST survey.}
\label{fig:sfs}
\end{center}
\end{figure*}

\begin{deluxetable*}{c|c|c|c}
\tablecolumns{4}
\tablecaption{Coefficients to the broken power-law parameterization (Equation \ref{eqn:bpl}) of the ridge and mean of the star-forming sequence\label{table:sfs_coeffs}. The uncertainties are calculated from Poisson uncertainties in the underlying density field.}
\tablehead{
\colhead{Parameter} & \colhead{$X_{0}$} & \colhead{$X_{1}$} & \colhead{$X_{2}$}
}
\startdata
\sidehead{Ridge (mode)}
\hline
$a$ & $0.03746 \pm 0.01739$ & $0.3448\pm0.0297$ & $-0.1156\pm0.0107$ \\
$b$ & $0.9605\pm0.0100$ & $0.04990\pm0.01518$ & $-0.05984\pm0.00482$ \\
$c$ & $0.2516\pm0.0082$ & $1.118\pm0.013$ & $-0.2006\pm0.0039$ \\
$\log M_t$ & $10.22\pm0.01$ & $0.3826\pm0.0188$ & $-0.04491\pm0.00613$ \\
\hline
\sidehead{\centering Mean}
\hline
$a$ & $-0.06707\pm0.00821$ & $0.3684\pm0.0128$ & $-0.1047\pm0.0044$ \\
$b$ & $0.8552\pm0.0068$ & $-0.1010\pm0.0107$ & $-0.001816\pm0.003290$ \\
$c$ & $0.2148\pm0.0045$ & $0.8137\pm0.0069$ & $-0.08052\pm0.00219$ \\
$\log M_t$ & $10.29\pm0.01$ & $-0.1284\pm0.0135$ & $0.1203\pm0.0044$ \\
\enddata
\end{deluxetable*}

The density $\rho(\log \mathrm{M}^*, \log {\rm SFR}, z)$ at several redshifts is shown in Figure \ref{fig:sfs}. The star-forming sequence is clearly visible as a peak in the density field. As stellar mass increases, the peak of the density increases and the distribution in log\,SFR grows wider. The overall increase in star-formation activity with increasing redshift can be seen clearly by examining the galaxy density. Additionally, the growth of the quiescent sequence can be seen, first appearing at high stellar masses then at later time extending to lower stellar masses as well.

We measure the evolution of the ridge (i.e. the number density peak) and the mean SFR as a function of mass and redshift. No star-forming galaxy selection is performed, i.e. this calculation is applied to the full density field. The ridge is measured as the peak of the conditional density field $\rho(\log {\rm SFR} \,| \log \mathrm{M}^*, z)$. In practice, the ridge is measured by starting at the stellar mass-complete limit and stepping upwards in increments of $\approx 0.01$ dex, requiring $|{\rm d}(\log {\rm SFR}) / {\rm d} (\log \mathrm{M}^*)| < 3$ at each step to ensure local continuity. This local continuity requirement prevents the ridge measurement from being affected by noise in the density field and also helps to avoid erroneously tracing the quiescent sequence where it dominates the density field. Starting at the low-mass end where the star-forming sequence is clearly defined, rather than the massive end where the density peak is lower and the shot noise is relatively high, helps to ensure the robustness of the measurement. While in principle there's no {\it a priori reason that the density will peak in the star-forming galaxy regime, particularly if quiescent galaxies dominate at the given mass and redshift, as we will demonstrate shortly the local continuity requirement ensures that this measurement is robust {\it a posteriori}.}
\begin{figure*}[ht]
\begin{center}
\includegraphics[width=0.95\linewidth]{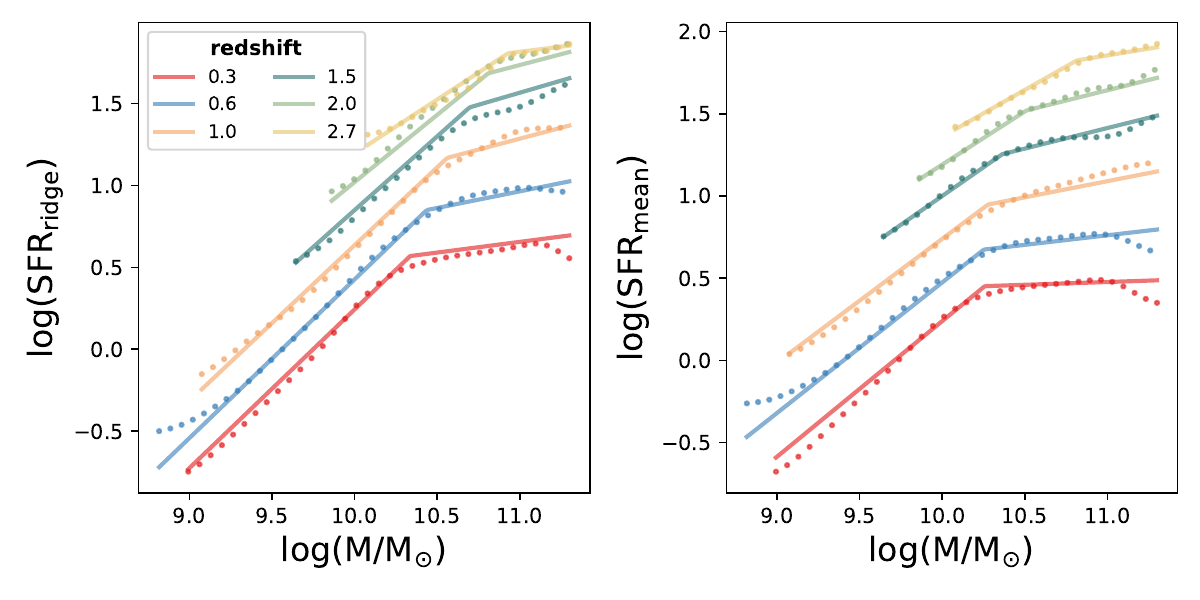}
\caption{The left panel shows evolution of the ridge of the star-forming sequence while the right panel shows the evolution of the mean star formation rate. The mean SFR includes contributions from both star-forming and quiescent galaxies. The points are measurements from the density field evaluated using the normalizing flow. The lines show the broken power-law fits described in Equations \ref{eqn:bpl} and \ref{eqn:ridge}.}
\label{fig:sfs_moments}
\end{center}
\end{figure*}

The goal of the nonparametric analysis is to avoid the bias created by assumption of a parametric form; however, it is also useful to create a compact form of the results to compare with the literature. Accordingly, the evolution of these relationships is parameterized using a double power-law in M$^*$ \citep{whitaker14,leja15}:
\begin{equation}
\label{eqn:bpl}
 \log(\mathrm{SFR}) = \left\{ 
  \begin{array}{l l}
     a(M - M_t) + c & \; \; \; \; \; M > M_t\\
     b(M - M_t) + c & \; \; \; \; \; M \le M_t
  \end{array} \right.
\end{equation}
where $M \equiv \log \mathrm{M}^*/\mathrm{M}_{\odot}$, $a$ is the slope at high masses, $b$ is the slope at low masses, $c$ is the y-axis intercept, and $M_t$ is the logarithm of the stellar mass where the slope transitions from the low-mass to the high-mass component.

The redshift evolution of each of these parameters is in turn parameterized by a quadratic in redshift. Linear and third-order polynomial fits were both explored, but the linear fit was found to have very poor predictive power and the third-order fit was found to be overly flexible without significantly improving the agreement with the density field. The equations describing the coefficients over $0.2 < z < 3.0$ follow the form
\begin{equation}
\label{eqn:ridge}
y_i= X_{i,0} + X_{i,1} z + X_{i,2} z^2
\end{equation}
where $y_i$ represents one of the four parameters in Equation \ref{eqn:bpl}. The equations describing the evolution in $\log(\mathrm{SFR}_{\mathrm{mean}})$ are defined in an analogous fashion. The mean and ridge are smoothed before fitting with Gaussian kernels with width $\delta z$ = 0.1, $\delta \log \mathrm{M}^*$ = 0.1 to remove small inhomogeneities in the density field. The coefficients are shown in Table \ref{table:sfs_coeffs}. The uncertainties are from counting (Poisson) uncertainties, i.e. underlying uncertainties in the density from the neural network. While the neural network produces only the maximum likelihood model as the output, it is conditioned using the uncertainties on stellar mass and star formation rate from SED-fitting, so these are also taken into account in this step. We note that the largest uncertainties are likely systematic errors endemic to SED-ftting. These have been estimated in a number of cross-code and simulation-based studies to be approximately 0.1 dex for stellar mass and 0.3 dex in SFR \citep{wuyts11a,wuyts11b,pforr12,mobasher15,carnall19,leja19a,lower20}.

The redshift evolution of the mean and ridge are shown in Figure \ref{fig:sfs_moments}. Both the parametric fit and the direct measurement from the density field are shown. The mean and the ridge increase steadily with increasing redshift up to $z\sim2$; above this, the mean continues to increase -- perhaps driven by continuing decrease in the fraction of passive galaxies -- whereas the evolution of the ridge begins to saturate. This is consistent with the observed flattening of sSFR evolution above $z\gtrsim 2.5$ (e.g., \citealt{santini17,leslie20}). Despite the fact that passive galaxies decrease the normalization of the mean SFR, the mean is still generally higher than the ridge (i.e. the peak of the density field), which is a standard feature in log-normal distributions such as the star-forming sequence.

\begin{figure*}[ht]
\begin{center}
\includegraphics[width=0.95\linewidth]{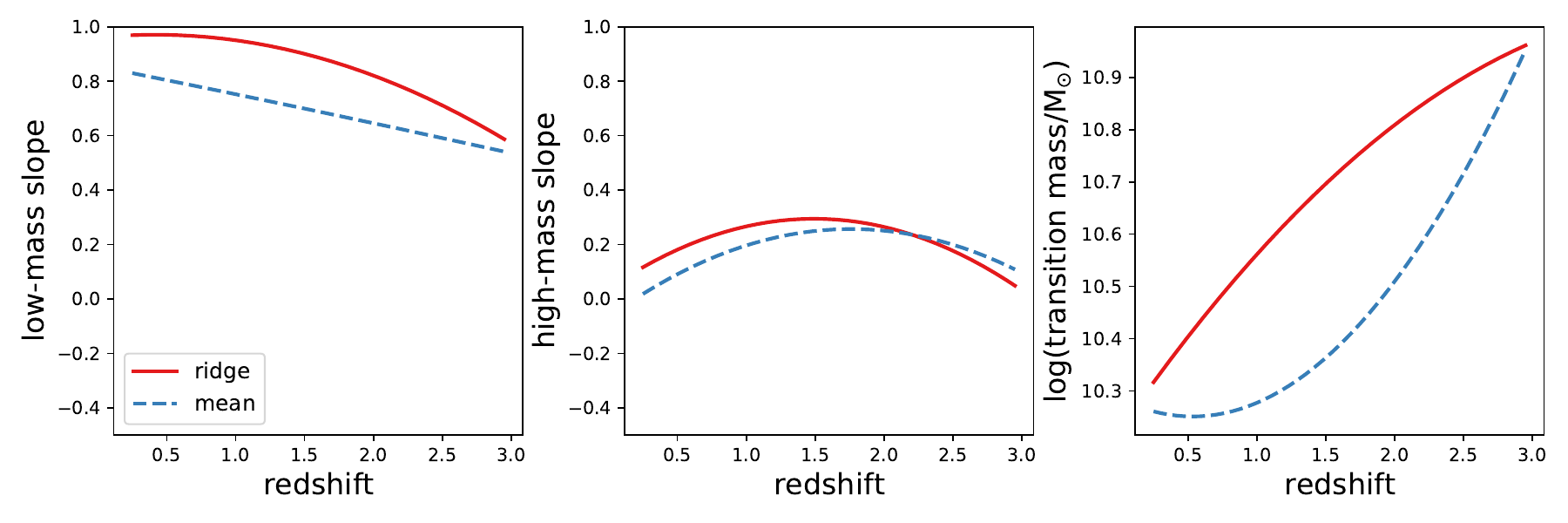}
\caption{The redshift evolution of the low- and high-mass slope and the stellar mass at which the slope transitions between the two. The high- and low-mass slopes are distinct at all redshifts. However, the need for multiple slopes in the ridge of the star-forming sequence is most clear at low redshifts. This is because at higher redshifts the transition mass increases to higher values of stellar mass, at which there are few ($N\sim5-10$, c.f. Fig. 1 in \citealt{leja20}) observable galaxies in these fields, meaning both the mean and ridge-line star formation rates are more weakly constrained.}
\label{fig:slopes_and_mt}
\end{center}
\end{figure*}

Figure \ref{fig:slopes_and_mt} shows the redshift evolution of the low-mass slope, high-mass slope, and transition mass. The low-mass slope of the ridge of the star-forming sequence stays near unity across most of the observed redshift range. Conversely, the high-mass slope stays relatively constant between 0 and 0.2, in contrast to high-mass slopes of 0.5-0.7 in found at $1 < z < 2.5$ in previous work but in agreement at $0.5 < z < 1$ \citep{whitaker14}. The transition mass increases with redshift, from log$M_t \sim 10.2$ at $z=0.2$ to logM$_t=11$ at $z=3$, at which point the high-mass slope is poorly constrained and even the existence of a two-slope star-forming sequence is not clear, due to the small number of objects at high masses and redshifts. The quadratic fit may underestimate the extent to which the high-mass slope is flat or even negative at high masses and low redshifts, as the density field in Figure \ref{fig:sfs} illustrates. This cannot be due to inaccuracies in the separation of star-forming/quiescent galaxies, since none is performed; instead, it suggests that there is little or no relationship between star formation rate and mass for galaxies with log(M/M$_{\odot}$)$\gtrsim 10.5$ between $0.2 < z \lesssim 1.5$. 

The low-mass slope of the mean behaves similarly to the ridge, though the slope is generally shallower at all redshifts. This is likely because the increase of quiescent galaxies with increasing stellar mass will naturally flatten out the average relationship between SFR and mass. The high-mass slope shows similar behavior. The transition mass for the mean is lower than for the ridge by $\sim0.1-0.5$ dex, likely reflecting the effect of an increasing quiescent fraction flattening out the slope at intermediate masses.

At the lowest masses there are small upturns in the SFR-M$^*$ relationship at some redshifts. This likely represents imperfections in the low-mass 90\% completeness limit, specifically a combination of up-scattering of bright galaxies with high sSFRs below the limit and an incomplete census of low-sSFR galaxies above the limit. The broken power-law fits are relatively robust to these effects and so no adjustment is made for them.

Finally, we note that there are deviations at the $\lesssim 0.1$ dex level between the broken power-law parameterizations and the true evolution of the mean and ridge-line of the galaxy star formation rate distribution. Such deviations are inevitable when fitting smooth functions to data with complex behaviors. For work with higher accuracy requirements, it is recommended to work directly on the trained density field, which is available online.

\subsection{The profile of the star-forming sequence}
\label{sec:derived_pars}
\begin{figure}[ht]
\begin{center}
\includegraphics[width=\linewidth]{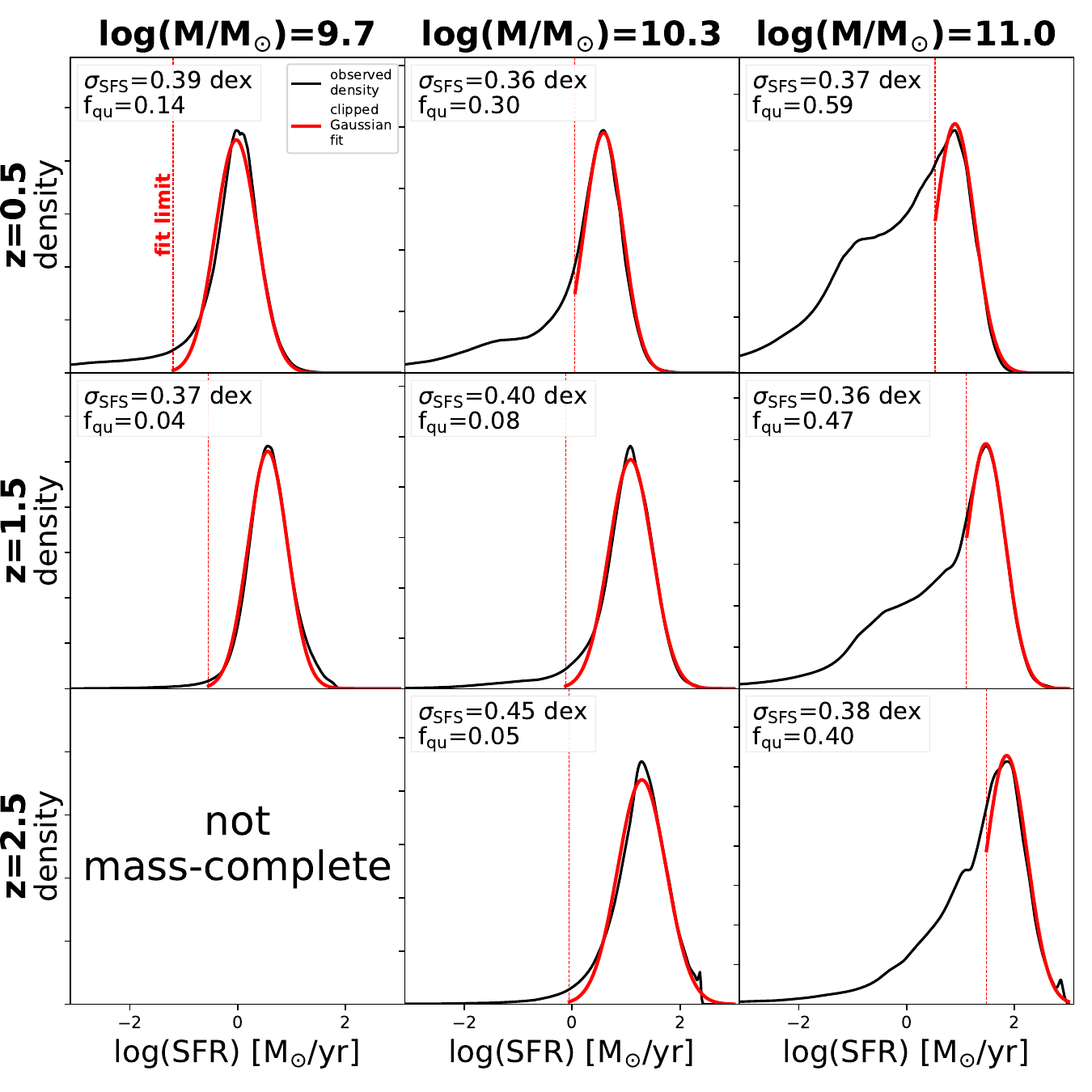}
\caption{Slices of the log\,SFR distribution at fixed values of log\,M$^*$ and redshift from the density estimate are in black, while Gaussian fits to the star-forming sequence shown in red. The fit is performed iteratively using sigma-clipping with a variable lower bound as described in the text. The fit limit in each panel is indicated with a dashed line. The stellar mass and redshift of each panel is indicated in the upper-left, along with the inferred width of the star-forming sequence. There is a fundamental trade-off between the fidelity of the Gaussian fit to the high-SFR end of the distribution and the exclusion of the long tail of quenching or quiescent galaxies below the star-forming sequence.}
\label{fig:sfs_slices}
\end{center}
\end{figure}
A Gaussian distribution is fit to the density, P(logSFR | logM$^*$, z), to determine the width of the star-forming sequence and the fraction of galaxies on this sequence. This is intended to produce a qualitative description of the evolving properties of the star-forming sequence population; detailed quantitative comparisons with other results in the literature are best performed using the ridge as inferred in the previous section, or using the density field directly.

Specifically, a Gaussian is fit to the distribution of $\rho(\log {\rm SFR} \,| \log {\rm M}^*, z)$, with the center of the Gaussian fixed to the ridge-line derived in Section \ref{sec:sfs_results}. A key challenge is that the distribution of the star-forming sequence looks increasingly less like a Gaussian distribution as the fraction of quiescent galaxies increases. This is addressed by performing an asymmetric sigma-clipping fit, with the clipping occurring {\it only} below the star-forming sequence. 

The point where the sigma-clipping process begins is chosen carefully to both accurately describe the density profile above the star-forming sequence while also excluding the long tail of quiescent galaxies below it. For each small slice in mass and redshift, an iterative sigma-clipping fit is performed to the density in log\,SFR with $\sigma_{\mathrm{clip}} = -3.0$. Here the negative sign indicates that clipping is performed to remove galaxies {\it below} the star-forming sequence. To ensure a high quality of fit, the fraction of galaxies existing {\it above} the ridge of the star-forming sequence is estimated from the fitted Gaussian. This is then compared to this same quantity estimated directly using the density from the trained normalizing flow. If these two estimates differ by more than 10\%, $\sigma_{\mathrm{clip}}$ is adjusted upwards by 0.5 in order to exclude more quiescent galaxies and the fit is performed again. We allow $\sigma_{\mathrm{clip}}$ to vary between $-3.0 \leq \sigma_{\mathrm{clip}} \leq - 1.0$. Examples of the fit to $\rho(\log {\rm SFR} \,| \log {\rm M}^*, z)$ are shown in Figure \ref{fig:sfs_slices}.

\begin{figure*}[ht]
\begin{center}
\includegraphics[width=0.95\linewidth]{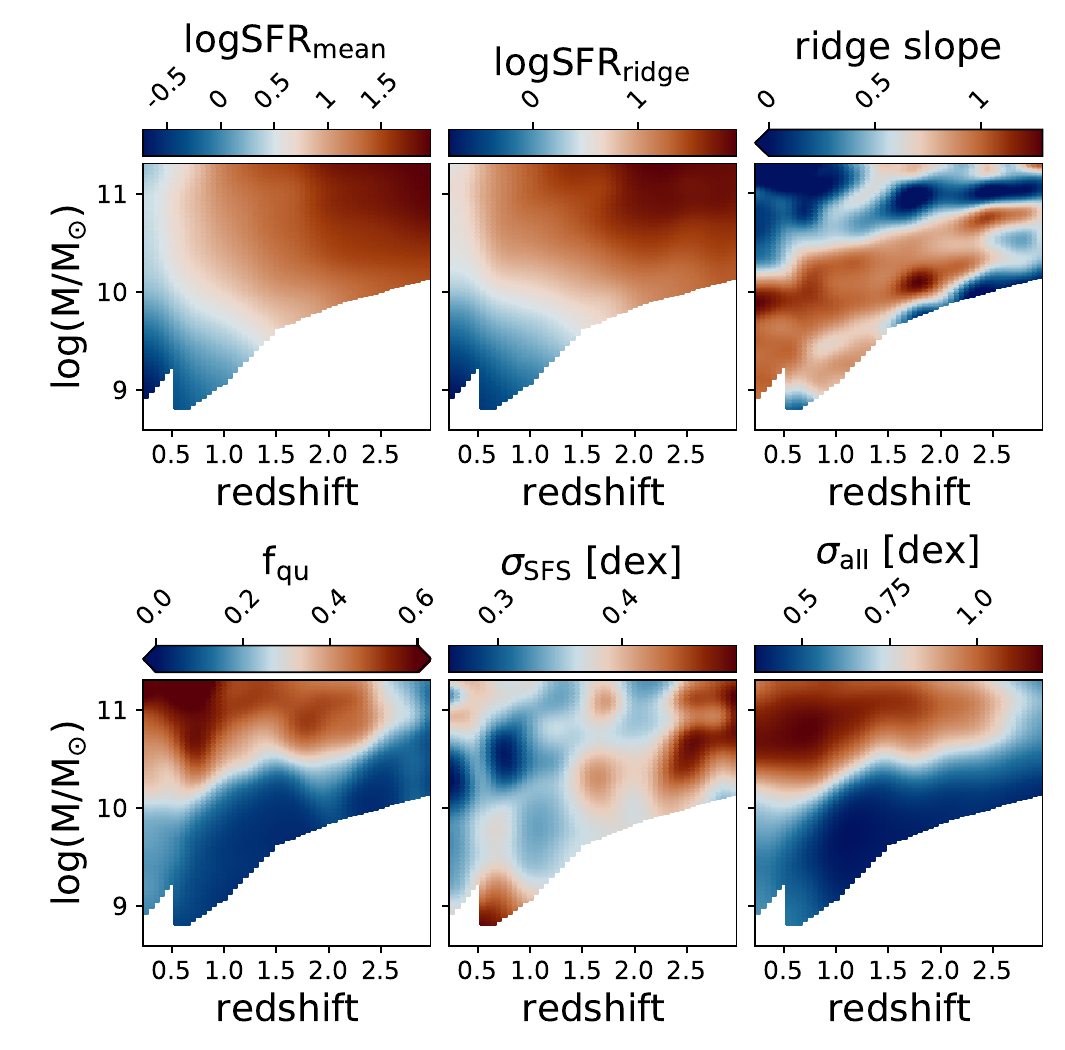}
\caption{The evolution of parameters derived from $\rho(\log {\rm SFR} \,| \log {\rm M}^*, z)$. In order of left to right, the first row shows the mean SFR, the ridge of the star-forming sequence, and the slope of the ridge, while the second row shows the quiescent fraction, the $1\sigma$ width of the star-forming sequence in logSFR, and the $1\sigma$ width in logSFR of the full population. The slope of the ridge is clipped at 0 to avoid having the color scale dominated by the down-turn at high masses and low redshifts. The parameter maps are smoothed with Gaussian kernels with width $\delta z$ = 0.1, $\delta$log\,M$^*$ = 0.1 for presentation purposes.}
\label{fig:derived_pars}
\end{center}
\end{figure*}

The redshift evolution of all parameters inferred from the density field are shown in Figure \ref{fig:derived_pars}. While the evolution of the ridge, mean, and slopes are parameterized explicitly in Equation \ref{eqn:bpl}, here we show the raw values as measured from the data. The width of the star-forming sequence comes from the fitted Gaussian, while the quiescent fraction is defined as (1 - fractional area covered by the fitted Gaussian). The parameter maps are smoothed with Gaussian kernels with width $\delta z$ = 0.1, $\delta$log\,M$^*$ = 0.1 for presentation purposes.

The maps of the slope and normalization (both mean and mode) of the star-forming sequence show evolution consistent with the parameterized fits in Figure \ref{fig:slopes_and_mt}. The quiescent fraction shows nearly monotonic increases with decreasing redshift and with increasing stellar mass, as expected. The width of the star-forming sequence from the log-normal fit is between $0.3-0.4$ dex at most masses and redshifts, increasing to $\sim$0.5 dex at higher masses where the distinction between star-forming and quiescent galaxies becomes blurred. Intriguingly, the width of the star-forming sequence also increases at high redshift, approaching $\sigma\sim0.5$ dex at $z>2$ for intermediate-mass galaxies.

Notably, the width of the star-forming sequence and the quiescent fraction are correlated with one another in the Gaussian profile and thus partially degenerate. Without strong pre-selection of star-forming galaxies and in the presence of a significant quiescent population, the width of the star-forming sequence is generally not very well constrained (e.g., Figure \ref{fig:bimodal_split}). The simple sigma-clipping procedure which produces the quiescent fractions and the width of the star-forming sequence in Figure \ref{fig:derived_pars} is only precise to within $\sim0.1$ dex (in width) or $\sim20\%$ (in quiescent fraction), and we recommend using the density from the trained flow directly for operations requiring higher precision than this. Given these caveats, the primary trend that we observe in the width of the star-forming sequence is that it is approximately constant between $0.3-0.4$ dex at $z<2$ and then becomes $0.4-0.5$ dex at $2 < z < 3$. Indeed, Fig. \ref{fig:bimodal_split} cleanly illustrates how the width of the star-forming sequence, $\sigma_{SFS}$, follows from the definition of a star-forming galaxy.

In contrast, the width of the full sequence (i.e., $\sigma_{all}$ = half of the 84$^{\mathrm{th}}$ - 16$^{\mathrm{th}}$ percentiles of the distribution) is independent of any distinction between star-forming and quiescent galaxies. This makes it a relatively robust property of the distribution, albeit at the `cost` of now being sensitive to the SFRs of non star-forming galaxies. At low masses and/or high redshifts where quiescent galaxies are rare, it effectively collapses to $\sigma_{SFS}$.

\section{Comparing to Previous Results and Techniques}
\label{sec:comparison}
This section compares the results from this work with previous results in the literature. Section \ref{sec:lit_comp} compares the star-forming sequence derived in this work to previous observational results, showing particularly that the normalization is considerably lower than previous work. Section \ref{sec:sfr_diff} explores the origin of this difference and introduces a general framework to interpret the difference between SED SFRs and standard UV+IR SFRs.

\subsection{A comparison with other star-forming sequences from the literature}
\label{sec:lit_comp}
\begin{figure*}[ht]
\begin{center}
    \includegraphics[width=0.95\linewidth]{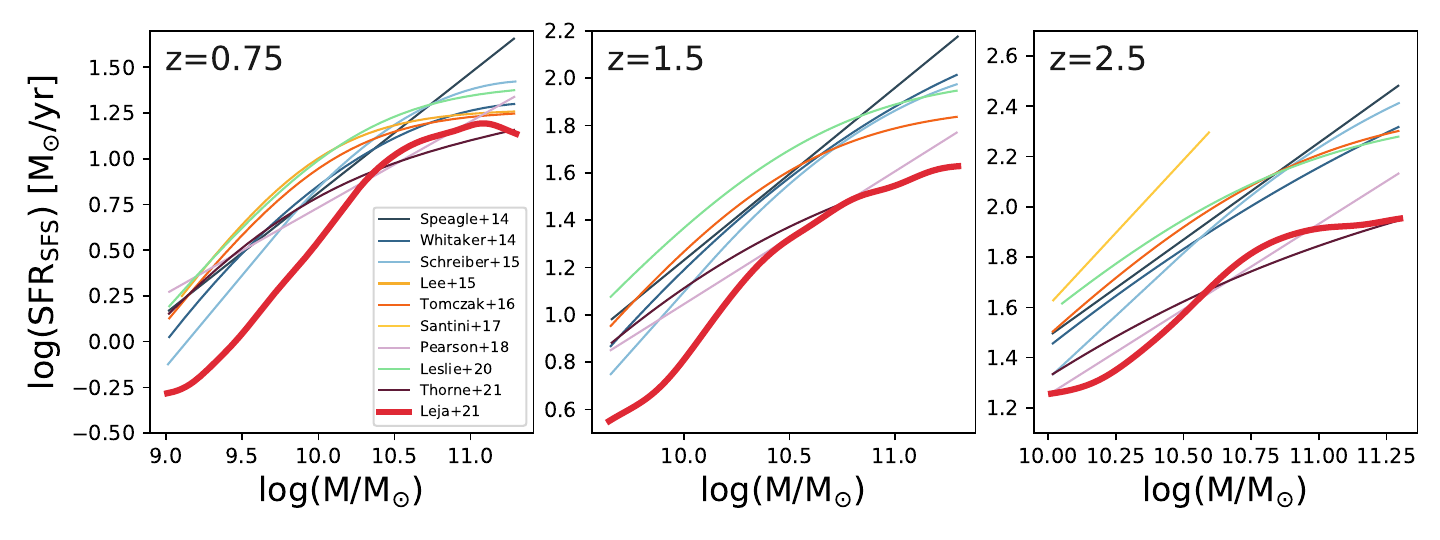}
\caption{Comparing the star-forming sequence derived in this work with other results from the literature. In general, the star-forming sequence inferred here has a normalization well below previous results. This is due to the methodology by which the stellar masses and star formation rates are inferred -- see Section \ref{sec:sfr_diff} for further discussion.}
\label{fig:literature}
\end{center}
\end{figure*}

Figure \ref{fig:literature} compares the star-forming sequence derived in this work to results in the literature \citep{speagle14,whitaker14,schreiber15,lee15,tomczak16,santini17,boogaard18,iyer18,pearson18,leslie20,thorne21}. The literature results are corrected to a Chabrier IMF to match this work, and all studies are mass-complete in this regime. The \citet{santini17} results are clipped at high masses where their assumption of linearity in the SFR-M relationship breaks down.

The star-forming sequence in this work is systematically lower than the measurements in the literature by $\sim0.2-0.5$ dex, with the offset decreasing at lower redshifts. This is not due to differences in definitions of the star-forming sequence, e.g. via pre-selection of star-forming galaxies. Such differences mostly affect the massive end (Figure \ref{fig:bimodal_split}), where the separation between star-forming and quiescent systems is the most confused (Figure \ref{fig:sfs_slices}). Instead, this offset is seen across the full range of stellar masses. This suggests the offset is instead due to differences in the inferred galaxy masses and star formation rates. These differences are explained in detail in the next section.

There are two notable exceptions to this systematic offset in the literature, specifically the studies of \citet{boogaard18} and \citet{iyer18}. What these two share in common is that they do not attempt to measure SFRs directly from broadband photometry: \citet{boogaard18} uses H$\alpha$ and H$\beta$ emission lines from the MUSE Hubble Ultra Deep Field, permitting an internal dust correction, and \citet{iyer18} infers the star-forming sequence slope and normalization indirectly, by constraining them with star formation histories derived from galaxies at lower redshifts. It is a positive sign that these relatively independent methodologies agree with the \prospector{}-$\alpha$ modeling of the broadband photometry.

Of the studies based on broadband photometry, the \citet{pearson18} and \citet{thorne21} results are most similar to the \prospector{}-$\alpha$ inferences. This is likely because they are also measuring both SFRs and stellar masses at the same time using SED-fitting. Solving for the stellar populations on a galaxy-by-galaxy basis naturally lowers the star formation rate estimates as compared to standard UV+IR formulae, as explained in the next section. Additionally, \citet{thorne21} also find more massive galaxies than in previous work; they similarly have chosen an SFH model which also returns older ages and thus higher stellar masses \citep{robotham20}. Yet the \prospector{}-$\alpha$ results still typically lie $\sim0.1-0.3$ dex below these inferences, with the offset increasing at low masses.

\subsection{The origin of the differences in the star-forming sequence}
\label{sec:sfr_diff}
\begin{figure*}[ht]
\begin{center}
    \includegraphics[width=0.95\linewidth]{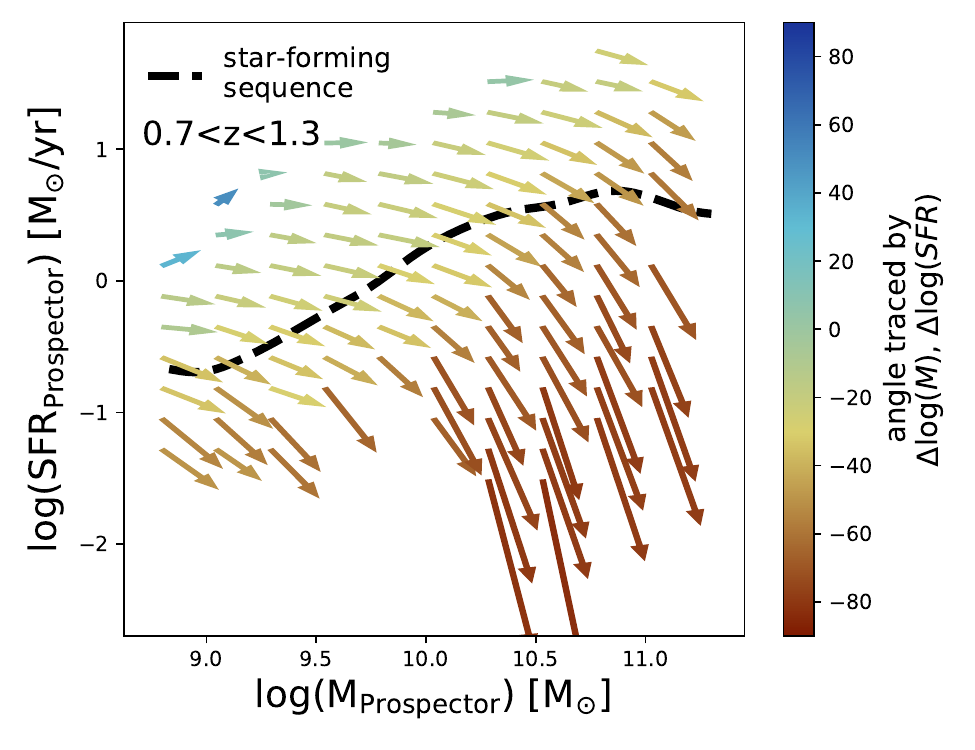}
\caption{The median shift in inferred values of log\,M$*$ and log\,SFR between the \prospector{}-$\alpha$ model and FAST stellar masses / UV+IR SFRs for galaxies at $0.7 < z < 1.3$. Only cells with $N>20$ galaxies are shown. The masses increase by $\sim0.2$ dex due to the use of nonparametric star formation histories. The inferred SFRs decrease by $0.1-1+$ dex largely due to dust heating from old stars, though other effects can contribute substantially in certain regimes (see Section \ref{sec:sfr_diff} for in-depth discussion). The ridge-line of the star-forming sequence as measured in this work is shown as a black dashed line. The color-bar traces the angle of the shift in logM-logSFR space, where $\theta=0$ is a horizontal arrow pointing to the right.}
\label{fig:deltam_deltasfr}
\end{center}
\end{figure*}

Here we explore the shift in inferred stellar masses and, specifically, star formation rates. We adopt the results from previous analysis of the 3D-HST survey as a reference point: stellar masses inferred using the FAST SED-fitting code \citep{kriek09} from \citet{skelton14}, and UV+IR SFRs from \citet{whitaker14}. Figure \ref{fig:deltam_deltasfr} shows the median shift in inferred values of log\,M$^*$ and log\,SFR between the \prospector{}-$\alpha$ values and the aforementioned works. The ridge of the star-forming sequence is included for reference.

The systematic increase in stellar masses is at this point well-understood to come from the use of nonparametric star formation histories, which find significantly older stellar ages (and thus higher stellar masses) than parametric methods. In brief, this is because they assume a larger fraction of old stars for a given observed SED; see \citet{carnall19,leja19a} for more details. 

There exists some evidence that these larger stellar masses are more accurate. Simple recovery tests using a wide range of input star formation histories show that common parametric star formation histories systematically under-estimate stellar ages \citep{carnall19,leja19a}. \citet{leja19b} showed that the more extended star formation histories inferred from nonparametric techniques are much more consistent with the observed evolution of the stellar mass function over $0.5 < z < 3$. \citet{lower20} tests these techniques on hydrodynamical simulations; they find that nonparametric techniques lead to significantly improved stellar masses in SED fitting, with an average bias of 0.4 dex with a delayed-$\tau$ model compared to a bias of less than 0.1 dex for a nonparametric model.

The difference in inferred SFRs has a more complex origin. \citet{leja19b} showed that the largest contribution to this shift is the effect of dust heating from `old' ($>$100 Myr) stars, which is incorrectly interpreted as star formation in standard UV+IR estimates. This effect is strongest for galaxies with low specific star formation rates, and can be seen in Figure \ref{fig:deltam_deltasfr} as a decrease in inferred SFRs for massive objects, with a magnitude increasing with decreasing SFR. However, this interpretation alone is incomplete: for example, it does not explain the increase in SFR for low-mass galaxies above the star-forming sequence. To address this more clearly, we generalize the UV+IR SFR equation to explain the physical origins of the difference between the \prospector{}$-\alpha$ SFRs and UV+IR SFRs. 

\subsubsection{The difference between UV+IR and SED SFRs}
The simple assumption underlying UV+IR SFRs is that the SFR is directly proportional to the luminosity of young stars, which follows from energy conservation:
\begin{equation*}
    \mathrm{SFR} = \alpha L_{\mathrm{bolometric,young\; stars}}
\end{equation*}
where $\alpha$ is a constant parameterizing the total energy output of young stars. This direct equivalence is why UV+IR SFRs are generally considered to be a reliable SFR estimator (e.g. \citealt{kennicutt98,kennicutt12}). One challenge in estimating $L_{\mathrm{bolometric,young\; stars}}$ is that young stars are often enshrouded in dust which attenuates much of their light and re-emits it in the infrared. The above equation then becomes
\begin{equation*}
    \mathrm{SFR} = \alpha(L_{\mathrm{unattenuated}} + L_{\mathrm{attenuated}})
\end{equation*}
Typically, these two components are estimated from observations by defining specific bandpasses in the UV and the IR and inferring the total energy emitted in these bands. This luminosity is then converted into a star formation rate by comparing to simple models of galaxy stellar populations. We use the templates from \citet{bell05} as an example. This work assumes $L_{\mathrm{unattenuated}} \propto L_{\mathrm{UV}}$ (1216 \AA$-3000$ \AA) and $L_{\mathrm{attenuated}} = L_{\mathrm{IR}}$(8$\mu$m$-1000\mu$m). They postulate a relation of the form
\begin{equation}
    \label{eqn:uvir}
    \mathrm{SFR}\; [M_{\odot}/yr] = \alpha(\delta L_{\mathrm{IR}} + \gamma L_{\mathrm{UV}}).
\end{equation}
\citet{bell05} compute the coefficients using a theoretical stellar population with a constant SFR over 100 Myr, finding coefficients (adjusted to a \citealt{chabrier03} IMF) $\alpha_0 = 1.09 \times 10^{-10}$ [M$_{\odot}$ yr$^{-1}$ L$_{\odot}^{-1}$], $\delta_0=1$, and $\gamma_0 = 2.2$. These values correspond to the parameters in the above equation, but are given subscripts in order to differentiate them from Prospector-inferred values for $\alpha$, $\delta$, and $\gamma$ introduced later in this section.

While the general form of Equation \ref{eqn:uvir} follows directly from energy conservation, the {\it coefficients} must be derived using stellar evolutionary theory and simplified models of the stellar populations in galaxies and can thus be subject to significant uncertainty. Standard UV+IR SFRs will assume simple model stellar population's properties to derive these coefficients (e.g., \citealt{bell05}), or infer them via empirical methods (e.g., \citealt{hao11}). Conversely, SFRs from SED-models such as \prospector{}-$\alpha$ will (implicitly) infer these coefficients on an object-by-object basis using models for the stellar populations in each galaxy. Notably, Equation \ref{eqn:uvir} itself holds true independent from the standard assumption of energy balance (i.e., $L_{\mathrm{attenuated}} = L_{\mathrm{emitted,IR}}$, \citealt{dacunha08}). Rather, energy balance is used in order to derive estimates of the coefficients $\alpha$, $\delta$, and $\gamma$.
\begin{figure*}[ht]
\begin{center}
    \includegraphics[width=0.95\linewidth]{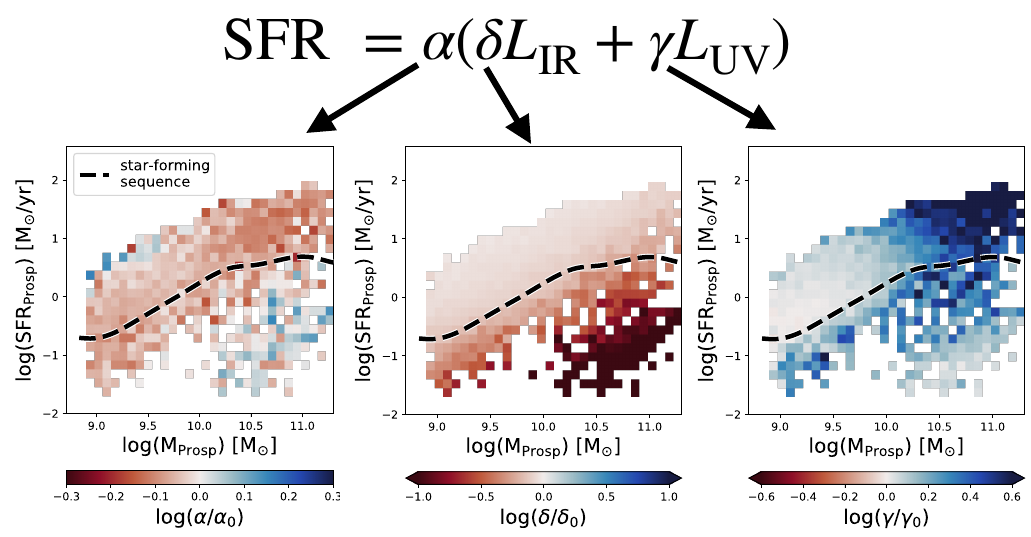}
\caption{The median difference in the coefficients to the UV+IR SFR in Eqn. \ref{eqn:uvir} between \prospector{}-$\alpha$ model and the fixed template-based coefficients from \citet{bell05} is shown for galaxies at $0.7 < z < 1.3$. $\alpha$ parameterizes the total energy output of young stars, $\delta$ parameterizes the fraction of IR luminosity which is powered by energy from young stars, and $\gamma$ parameterizes the conversion from the observed UV luminosity to the bolometric luminosity of the young stars {\it after} dust attenuation.  Arrows on the colorbar indicate where the data is clipped to enhance the information content of the map. The y-axis shows the SFR averaged over the most recent 100 Myr}.
\label{fig:uvir_coeffs}
\end{center}
\end{figure*}

We directly calculate the UV+IR coefficients for each galaxy from the \prospector{}-$\alpha$ physical models by perturbing the star formation rates from their measured values and measuring the response in L$_{\mathrm{IR}}$ and L$_{\mathrm{UV}}$. Figure \ref{fig:uvir_coeffs} shows the median difference between the \prospector-$\alpha$ coefficients and those from the \citet{bell05} templates in the SFR-mass plane. These differences are discussed in turn below.

The first coefficient, $\alpha$, represents the total luminosity of young stars (`young' here defined as age $<100$ Myr). At these redshifts, \prospector{}-$\alpha$ finds median values in the SFR-M plane of $-0.2 \lesssim \log(\alpha_{\mathrm{Prosp}}/\alpha_0) \lesssim 0.15$, though the scatter can vary up to $\pm0.4$ for individual galaxies.

In the \prospector-$\alpha$ model, variations in $\alpha$ are primarily determined by the SFR(0-30 Myr)/SFR(30-100 Myr) ratio (Spearman-$r$ coefficient of 0.81). This is a straightforward consequence of the fact that younger stars (0-30 Myr) are relatively more luminous than older stars (30-100 Myr). Thus, galaxies with rising star formation histories have lower values of $\alpha$ and galaxies with falling star formation histories have higher values of $\alpha$. Since galaxies on and above the star-forming sequence tend to have rising star formation histories while galaxies below have falling star formation histories \citep{leja19b}, this naturally produces an increase in $\alpha$ with decreasing SFR at a fixed stellar mass. This relationship has considerable scatter though, particularly at the low-mass end, suggesting that some galaxies are being observed in the process of moving back towards their equilibrium values (i.e., around the ridge-line of the star-forming sequence).

The second coefficient, $\delta$, represents the fraction of the emitted IR luminosity which is powered by young stars. The primary other source of energy is old stellar populations, although AGN can be significant in a minority of cases. In the \citet{bell05} UV+IR formulation, $\delta=1$ because there is no other source of energy; however, it has long been known that $\delta$ {\it should} be $<1$ (e.g., \citealt{fumagalli14,utomo14,hayward15}). In empirical calibrations $\delta$ is inferred to be $\sim0.6$ for star-burst galaxies (e.g., \citealt{meurer99}), or $0.46$ for normal star-forming galaxies \citep{hao11}. 

\prospector{}-$\alpha$ infers $0.01 \lesssim \delta \leq 1$. Variations in $\delta$ in the \prospector{}-$\alpha$ model are almost entirely driven by heating from old stars: \citet{leja19b} finds a strong relationship between specific star formation rate (sSFR) and the fraction of dust heating performed by old stars, with galaxies at log(sSFR/yr$^{-1}) = -11$ having $\sim$80\% of their dust emission powered by old stars. This relationship is in excellent qualitative agreement with detailed radiative transfer simulations of nearby galaxies from Dustpedia \citep{nersesian20}. Notably, this effect gets {\it stronger} at higher redshifts, where the `old' stars are relatively more luminous than at later times. This relationship is evident in Figure \ref{fig:uvir_coeffs} as a decrease in $\delta$ with decreasing SFR at a fixed stellar mass, ranging from $\delta \approx 1$ for highly star-forming galaxies to $\delta << 1$ for quiescent galaxies.

The final coefficient, $\gamma$, represents the conversion from the observed UV luminosity at $1216-3000$ \AA\: to the bolometric luminosity of young stars. This measurement is performed {\it after} dust attenuation. Thus it represents the complement to the IR luminosity, ensuring that the full energy budget of young stars is included in the estimate.

Variations of $\gamma$ from \prospector{}-$\alpha$ span a range of $-0.05 \lesssim \log(\gamma/\gamma_0) \leq 1.1$. The $\gamma$ coefficient correlates with the shape of the SED of young stars after the effects of dust attenuation. The largest determinant of $\gamma$ in \prospector{}-$\alpha$ is perhaps unsurprisingly $\tau_{\mathrm{diffuse\;dust}}$ (Spearman-$r$ coefficient of 0.88). This is the normalization of the dust attenuation curve (closely related to $A_V$) and represents the most important determinant of the total amount of reddening in the observed SED. Stellar metallicity is an important secondary effect as decreasing stellar metallicity will make the SED of young stars bluer (Spearman-$r$ coefficient of 0.4). The $\gamma$ map in Figure \ref{fig:uvir_coeffs} generally increases with both mass and star formation rate, consistent with trends in the observed dust attenuation \citep{whitaker17}.

The sum of these effects can now explain the full transformation of the star-forming sequence in Figure \ref{fig:deltam_deltasfr}. Galaxies on and below the star-forming sequence have lower inferred SFRs due to the dust-heating effect from old stars. Galaxies above the star-forming sequence show little effect on their star formation rates; here, the standard assumption that L$_\mathrm{IR}$ is entirely powered by star formation appears to hold. Conversely, low-mass galaxies above the star-forming sequence show an increase in their SFRs due to a combination of the effects of dust on the interpretation of their UV luminosities and bursty recent star-formation histories. 

An important point is that the IR energy dominates the UV+IR equation above $\log$(M/M$_{\odot}) \gtrsim 9.5$ \citep{whitaker17}), while UV energy dominates for galaxies with $\log$(M/M$_{\odot}) \lesssim 9.5$; this means that $\delta$ will have a larger influence at high masses, and $\gamma$ a larger influence at low masses.

We emphasize that while measuring these coefficients with a model such as \prospector{}-$\alpha$ is a more robust approach than using fixed coefficients, these inferences are still based on simplified models of galaxy stellar populations and should be viewed with scrutiny. One method to investigate their accuracy is via posterior predictive checks, where one set of observations is fit by a model which is in turn used to predict a second independent set of observations. The star formation rates from \prospector{}-$\alpha$ perform well here, finding a scatter of 0.2 dex and negligible offset when fitting UV-IR broadband photometry and predicting measurements of H$\alpha$ \citep{leja17} and Br-$\gamma$ \citep{pasha20} emission line luminosities. Br-$\gamma$ predictions are particularly informative: Br-$\gamma$ is a hydrogen emission line at $\sim2.16\mu$m which is nearly insensitive to dust attenuation, thus serving as a dust-independent measure of star formation rates. However, these tests are complicated by the systematic $\sim0.2$ dex sensitivity of emission line luminosities to the chosen stellar isochrones and their treatment of rotation or binary effects \citep{steidel16,choi17,pasha20}. Additional posterior predictive checks using a wide range of observables across a wide range of galaxy types and redshifts will be beneficial in further calibrating SED model outputs.

\section{Discussion}
\label{sec:discussion}
Here the broader implications of the results are discussed. Section \ref{sec:theory} compares the star-forming sequence in this work with those from hydrodynamical simulations of galaxy formation, and Section \ref{sec:philosophy} discusses the prospects for robust measurements of the star-forming sequence in the absence of a clear star-forming/quiescent galaxy separation.

\subsection{A new agreement with theoretical models of galaxy formation}
\label{sec:theory}
There has been a long-standing systematic offset of $\sim0.2-0.5$ dex in the normalization of the $0.5 \lesssim z \lesssim 3$ star-forming sequence between observations and predictions from galaxy formation simulations  \citep{mitchell14,leja15,furlong15,tomczak16,donnari19,dave19,katsianis20}. This offset is such that simulations under-predict the observations. This offset is persistent across most of the major theoretical models of galaxy formation. The universality comes from the fact that this problem is challenging to solve with baryonic feedback prescriptions; galaxy star formation rates typically track the gas accretion rate, which is in turn set by the behavior of dark matter at large scales \citep{mitchell14}.

\begin{figure*}[ht]
\begin{center}
    \includegraphics[width=0.95\linewidth]{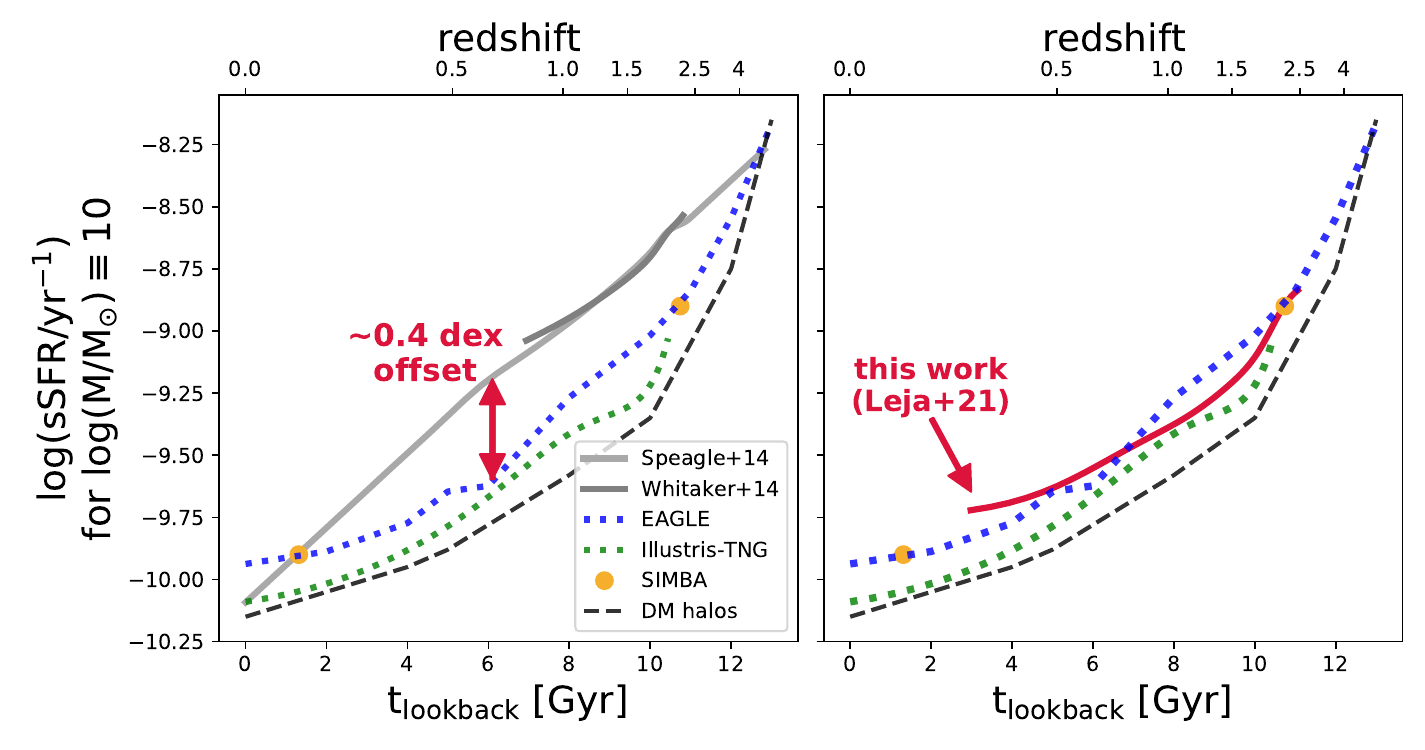}
\caption{This figure compares the redshift evolution of specific star formation rate for a log(M/M$_{\odot})=10$ galaxy on the star-forming sequence between observations and galaxy formation simulations. The specific dark matter accretion rate from the Millenium simulation is also shown for reference \citep{fakhouri10}. The left panel shows previous observational star-forming sequences, which have a $\sim0.4$ dex offset with the simulations. The right panel shows results from this work which resolve these systematic differences.}
\label{fig:ssfr_vs_dm}
\end{center}
\end{figure*}
We show that this tension is resolved by the star-forming sequence presented in this work (Figure \ref{fig:ssfr_vs_dm}), confirming preliminary work in \citet{nelson21}. This figure shows the redshift evolution of the specific star formation rate for a log(M/M$_{\odot})=10$ galaxy from the observations \citep{speagle14,whitaker14} and from this work. Predictions for the galaxy star-forming sequence are shown from several major cosmological hydrodynamical simulations of galaxy formation: Illustris-TNG \citep{donnari19}, the EAGLE simulations \citet{furlong15}, and SIMBA \citep{dave19}. The specific dark matter accretion rates are taken from the Millenium N-body simulations \citep{fakhouri10}. Whereas before there was a systematic offset of $\sim$0.2$-0.5$ dex with the observations, there is now broad agreement between the observations, the hydrodynamical simulations, and the dark matter accretion rates.

This new agreement is important to the field because it is now possible to use the observed star-forming sequence across a wide range of redshifts as a constraint on hydrodynamical simulations of galaxy formation. \citet{hahn19} measure the star-forming sequence across multiple simulations using a consistent nonparametric definition and show that the simulated star-forming sequences differ by up to 0.7 dex in normalization and have significantly different slopes, in spite of the fact that they all approximately match the $z=0$ galaxy stellar mass function \citep{somerville15}. Having the new capacity to adjust their physics and feedback models to match the $0.5 < z < 3$ star-forming sequence will help to bring consensus in models of galaxy formation.

Constraints on the star-forming sequence will in turn provide a strong constraint on the feedback models in simulations, which currently have substantial qualitative and quantitative differences (see the review in \citealt{somerville15} for discussion). We note that the release of tools like the trained normalizing flow used in this work means that these comparisons can now be done directly in the space of the SFR-M$^*$-redshift density, $\rho(\log {\rm SFR}, \log {\rm M}^*, z)$. This bypasses classic challenges in comparing observations and simulations, such as pre-selection of star-forming galaxies or definitions of the star-forming sequence.

\subsection{Is it meaningful to enforce a separation between star-forming and quiescent galaxies?}
\label{sec:philosophy}
It has long been known that galaxies have clear bimodal distributions in rest-frame color diagrams such as the red sequence \citep{bell04} or the UVJ diagram \citep{williams09}. This naturally leads to the idea of two separate classes of galaxies: star-forming and quiescent. In such a framework, it is straightforward to use colors or a SFR indicator such as H$\alpha$ to distinguish between the two classes. It is also natural to focus more interest on measuring the star formation rates in star-forming galaxies (i.e. the star-forming sequence), as growth by star formation in quiescent galaxies is thought to be negligible.

However, the distribution of galaxies in the SFR-mass plane inferred by this analysis -- importantly, with no pre-selection of star-forming galaxies -- instead resembles a unimodal distribution with a long skew or tail towards low star formation rates (e.g., Figure \ref{fig:sfs_slices}). \citet{feldmann17} observe a similar distribution, and argue that the star-forming/quiescent galaxy distinction as defined by a log-normal star-forming sequence is inconsistent with the idea of star formation as a discrete and burst-driven event. Detailed studies of the local universe such as the Herschel Reference Survey show no separation at all between star-forming and quiescent galaxies \citep{eales17}. In this new schema, using SFR indicators to distinguish between star-forming and quiescent galaxies is not a robust methodology due to the strong overlap between the two classes. This danger is compounded by the fact that SFR indicators at low sSFRs are (a) much noisier due to a lower intrinsic brightness, and (b) and more prone to be contaminated with emission from processes unrelated to star formation (e.g., LINER emission, AGN, or dust heating by old stars).

It may not yet be possible to quantitatively distinguish between these two schema, at least not using these data. This is because colors can be bimodal even if sSFR is not, as photometric colors saturate at low sSFRs (see e.g. \citealt{leja20} and references therein). Put another way, the key challenge is that the observed SFRs become highly uncertain at low sSFR. Figure \ref{fig:nsamp_comparison} provides a clear demonstration of this. The sensitivity of the star-forming sequence at low SFRs to the uncertainty resolution suggests that the inferred shape of the star-forming sequence at low sSFRs is not a robust measurement, but instead, largely traces the shape of the uncertainties in SFRs.

Regardless of whether there is truly a bimodal sequence that is unobservable because it is smeared out by currently-irreducible observational error, or whether there is indeed no bimodality at all, there are clear practical ramifications to this question. Figure \ref{fig:bimodal_split} demonstrates that the normalization and width of the star-forming sequence and the fraction of quiescent galaxies both depend strongly on the separation of star-forming and quiescent galaxies at the $\sim0.2-0.3$ dex level, even when using the same star formation rate indicator. Given this, it is unsurprising that the observed width of the star-forming sequence varies between $\sim0.25-0.5$ dex in the literature \citep{speagle14}.

In light of these uncertainties, we argue that it is more meaningful to compare measurements of the density in the SFR-mass plane than it is to parameterize the star-forming sequence itself using a bimodal galaxy classification system. The ridge-line of the star-forming sequence is largely invariant to bimodal classification \citep{renzini15}, and the $1\sigma$ scatter of the {\it full} galaxy population in SFR-M is also meaningful and straightforward to measure. In this work, we additionally package the inferred density of the star-forming sequence into a trained normalizing flow and make it publicly available. In this way, galaxy formation models and future observational studies can compare to the full set of observations rather than attempting to emulate fragile observational selection criteria. This is an important step on the path towards likelihood-free inference. This approach has the additional advantage of including the full structure in the SFR-stellar mass plane, including such features as star-bursting galaxies above the star-forming sequence and the distribution of `green valley' and quiescent galaxies below the star-forming sequence.

\section{Conclusion}
\label{sec:conclusion}
In this paper, we infer the galaxy star-forming sequence across $0.2 < z < 3$ using a combination of machine-learning techniques and nonparametric methods applied to state-of-the-art UV-IR galaxy SED modeling from \prospector-$\alpha$. Specifically, we train a normalizing flow to reproduce the distribution $\rho(\log {\rm M}^*, \log {\rm SFR}, z)$ of galaxies in log\,M$^*$, log\,SFR, and redshift, with alterations included to marginalize over the full log\,M$^*$ and log\,SFR uncertainties derived from our previous \prospector{}-$\alpha$ fits.

%The resulting star-forming sequence has a lower normalization than nearly every previous result in the literature. We argue that this lower normalization is a natural result of the increased sophistication in our physical model for galaxy SEDs, and demonstrate that this lower normalization brings new-found agreement with theoretical models of galaxy formation.

Our conclusions are summarized here:
\begin{itemize}
    \item We use the \prospector{}-$\alpha$ fits to show that pre-selection of star-forming galaxies adds substantial systematic uncertainty to the inferred properties of the star-forming sequence. Applied to the same data, four standard but different pre-selection methods methods produce differences of $\sim$0.3 dex in the normalization, a $\sim$0.2 dex in the width, and $\sim$20\% differences in the fraction of quiescent galaxies.
    \item We instead present mass- and redshift-dependent fits to the ridge-line (number density peak) of the star-forming sequence and the mean SFR for all galaxies, avoiding fragile observational selection criteria for star-forming galaxies.
    \item We show that the star-forming sequence has a mass-dependent slope below $z\sim2$, with a flatter relationship at high masses and a steeper relationship at low masses with a slope near unity. Above $z\sim2$ a single slope is able to fit the data.
    \item The profile of the star-forming sequence is relatively narrow ($\sigma \sim 0.3-0.35$ dex) across a wide range of masses and redshifts. It is, however, significantly wider ($\sigma \sim0.4-0.5$ dex) at high redshift ($z \gtrsim 2$) and high masses (M$^* > 10^{11}$ M$_{\odot}$).
    \item The star-forming sequence inferred in this work is $0.2-0.5$ dex below the great majority of previous results in the literature. We show that this lower normalization is specifically caused by the nonparametric SFHs in \prospector{}-$\alpha$ which produce older ages and higher stellar masses and the overall lower inferred SFRs.
    \item A general framework is developed to compare SED-based SFRs with UV+IR SFRs.  This framework demonstrates that inferring star formation histories, dust attenuation, and metallicities on an object-by-object basis via SED-fitting produces more sophisticated (and, likely, more accurate) conversions between the observed flux and the SFRs. It is shown that the net effect of these changes in \prospector{}-$\alpha$ produce lower SFRs than standard UV+IR formulae, particularly at higher redshifts.
    \item Our new star-forming sequence follows the specific dark matter accretion rate much more closely than previous work, removing a long-standing $0.2-0.5$ dex offset at $0.5 < z < 3$ in the star-forming sequence between observations and predictions from models of galaxy formation.
    \item Our trained normalizing flow describing the full density $\rho(\log {\rm M}^*, \log {\rm SFR}, z)$ of galaxies in log\,M$^*$, log\,SFR, and redshift is made available online \href{https://github.com/jrleja/sfs_leja_trained_flow}. This permits direct comparison between this work and future theoretical and observational works, removing the need to define a star-forming sequence or separate star-forming and quiescent galaxies.
\end{itemize}

In this work, we have used state-of-the-art tools in machine learning, statistics, and computing to solve long-standing problems in observational galaxy evolution. In the future, we look forward to employing more sophisticated models and better tools to further address extant issues in this work. Some avenues for future improvement include better accounting for the effects of cosmic variance (sampling variance), more sophisticated modeling of the selection function near the stellar mass-complete limit, and a significant reduction of the computational resources required to fit high-dimensional SED models. 

Finally, the galaxy star-forming sequence is only as accurate as the models used to produce it. Due to the complex nature of stellar populations in galaxies, these physical models greatly benefit from cross-calibration tests to ensure their accuracy. Such tests for \prospector{}-$\alpha$ have produced promising results in spectroscopic observations of nearby galaxies \citep{leja17,pasha20}, in hydrodynamical simulations \citep{lower20}, and in comparisons between low-redshift star formation histories and high-redshift observations \citep{leja19b}. However, existing posterior predictive tests typically only cover small, well-studied samples or simulated objects (`models fitting models'). Going forward, there is a sore need for tests covering a variety of stellar populations parameters across a wide range of both stellar masses and cosmic times, particularly in the early universe, to ensure robust parameter inference.

\software{
\texttt{Astropy} \citep{astropy13,astropy18},
\texttt{dynesty} \citep{speagle20}
\texttt{FSPS} \citep{conroy09b}, 
\texttt{ipython} \citep{ipython}, 
\texttt{matplotlib} \citep{matplotlib18}, 
\texttt{numpy} \citep{numpy}, 
\texttt{Prospector} \citep{johnson21}, 
\texttt{python-fsps} \citep{pythonfsps14}, 
\texttt{PyTorch} \citep{torch19},
\texttt{scipy} \citep{virtanen20}
}

\acknowledgements We thank the anonymous referee for a thorough report which substantially improved the quality of the paper. This work is based on data products from observations made with ESO Telescopes at the La Silla Paranal Observatory under ESO program ID 179.A-2005 and on data products produced by TERAPIX and the Cambridge Astronomy Survey Unit on behalf of the UltraVISTA consortium. YST is grateful to be supported by the Australian Research Council DECRA Fellowship DE220101520 and the NASA Hubble Fellowship grant HST-HF2-51425.001 awarded by the Space Telescope Science Institute.

\bibliography{jrlbib}

\end{document}